\documentclass[twocolumn,prb,showpacs,superscriptaddress,preprintnumbers,amsmath,amssymb]{revtex4}  %Old one

\usepackage{graphicx}% Include figure files
\usepackage{dcolumn}% Align table columns on decimal point
\usepackage{bm}% bold math

\begin{document}

%\preprint{draft, Andrea Thomas}

\title{Crystal-field and Kondo scale investigation of CeMIn$_5$ (M=Co, Ir and Rh):\\
a combined x-ray absorption and inelastic neutron study}
\author{T. Willers}
  \affiliation{Institute of Physics II, University of Cologne,
   Z{\"u}lpicher Stra{\ss}e 77, D-50937 Cologne, Germany}
\author{Z. Hu}
% \altaffiliation [present address: ] {Max Planck Institute CPfS, N{\"o}thnizer Stra{\ss}e 40, 01187 Dresden, Germany}
 \affiliation{Institute of Physics II, University of Cologne,
   Z{\"u}lpicher Stra{\ss}e 77, D-50937 Cologne, Germany}  
 \affiliation{Max Planck Institute CPfS, N{\"o}thnizer Stra{\ss}e 40, 01187 Dresden, Germany}
\author{N. Hollmann}
  \affiliation{Institute of Physics II, University of Cologne,
   Z{\"u}lpicher Stra{\ss}e 77, D-50937 Cologne, Germany}
\author{P. O. K\"orner}
  \affiliation{Institute of Physics II, University of Cologne,
   Z{\"u}lpicher Stra{\ss}e 77, D-50937 Cologne, Germany}
\author{J. Gegner}
  \affiliation{Institute of Physics II, University of Cologne,
   Z{\"u}lpicher Stra{\ss}e 77, D-50937 Cologne, Germany}
\author{T. Burnus}
  \altaffiliation [present address: ] {IFF, Research Centre J{\"u}lich, 52425 J{\"u}lich, Germany}
  \affiliation{Institute of Physics II, University of Cologne,
   Z{\"u}lpicher Stra{\ss}e 77, D-50937 Cologne, Germany} 
\author{H. Fujiwara}
  \affiliation{Institute of Physics II, University of Cologne,
   Z{\"u}lpicher Stra{\ss}e 77, D-50937 Cologne, Germany} 
\author{A. Tanaka}
  \affiliation{Department of Quantum Matter, ADSM Hiroshima
  University, Higashi-Hiroshima 739-8530, Japan}
\author{D. Schmitz}
\affiliation{Helmholtz-Zentrum Berlin, BESSY II, Albert-Einstein-Stra{\ss}e 15, D-12489 Berlin,
  Germany}
\author{H. H. Hsieh}
  \affiliation{Chung Cheng Institute of Technology, National Defense University,
  Taoyuan 335, Taiwan}
\author{H.-J. Lin}
  \affiliation{National Synchrotron Radiation Research Center (NSRRC), 101 Hsin-Ann
  Road, Hsinchu 30077, Taiwan}
\author{C. T. Chen}
  \affiliation{National Synchrotron Radiation Research Center (NSRRC), 101 Hsin-Ann
  Road, Hsinchu 30077, Taiwan}
\author{E.D. Bauer}
	\affiliation{Los Alamos National Laboratory, Los Alamos, New Mexico 87545, USA}  
\author{J.L.Sarrao}
	\affiliation{Los Alamos National Laboratory, Los Alamos, New Mexico 87545, USA}  
\author{E. Goremychkin}
  \affiliation{ISIS, Rutherford Appleton Laboratory, Chilton, Didcot, Oxon, OX11 0QX, United Kingdom}
\author{M. Koza}
  \affiliation{Institut Laue Langevin, 6 rue Horowitz, 38042 Grenoble, France}
\author{L. H. Tjeng}
%  \altaffiliation [present address: ] {Max Planck Institute CPfS, N{\"o}thnizer Stra{\ss}e 40, 01187 Dresden, Germany}
  \affiliation{Institute of Physics II, University of Cologne,
   Z{\"u}lpicher Stra{\ss}e 77, D-50937 Cologne, Germany}
  \affiliation{Max Planck Institute CPfS, N{\"o}thnizer Stra{\ss}e 40, 01187 Dresden, Germany}
\author{A. Severing}
  \affiliation{Institute of Physics II, University of Cologne,
   Z{\"u}lpicher Stra{\ss}e 77, D-50937 Cologne, Germany}

\date{\today}

\begin{abstract}
Linear polarized soft-x ray absorption (XAS) and inelastic neutron scattering (INS)
experiments have been performed on CeMIn$_5$ with M = Rh, Ir, and Co  to determine 
the crystal-field scheme and characteristic Kondo temperatures T$^*$ 
for the hybridization between $4f$ and conduction electrons.
 The ground state wave functions are determined from the polarization dependent soft-XAS data 
 at the cerium M$_{4,5}$ edge and the crystal-field splittings from INS. The 
 characteristic temperature T$^*$ has been determined 
from the line widths of the neutron scattering data. We find that the quasielastic line widths 
of the superconducting compounds CeCoIn$_5$ and CeIrIn$_5$ are comparable with the low energy 
crystal-field splitting.

\end{abstract}


\maketitle

\section{Introduction}

The ternary rare earth family CeMIn$_5$ (M = Co, Ir, and Rh) are
heavy fermion, unconventional superconductors:
\cite{PetrovicSCinCeCoIn5,Sidorov2002} 
CeCoIn$_5$ and CeIrIn$_5$ become superconducting at ambient pressure 
at $T_c=2.3$\cite{PetrovicSCinCeCoIn5} 
and $T_c=0.4$ K\cite{PetrovicSCinCeIrIn5}, whereas 
the incommensurate heavy fermion antiferromagnet CeRhIn$_5$ ($T_N=3.8$) 
exhibits superconductivity under pressure ($P_c=1.6$ GPa, 
$T_c=2.1 K$).\cite{HeggerSCinCeRhIn5} All members of this family crystallize 
in the tetragonal HoCoGa$_5$ structure (space group P4/\textit{mmm}), 
which is derived from cubic CeIn$_3$ intercalated with MIn$_2$ layers 
along the tetragonal $c$ axis. The cubic compound CeIn$_3$ orders antiferromagnetically at
$T_N$ = 10 K and has a hybridization temperature $T^*$ of about 10 K.\cite{Knafo2003} 
This temperature, which gives the energy scale of the hybridization 
between the local $4f$ moments and surrounding conduction electrons, is 
highly pressure dependent in CeMIn$_5$ compounds. 
It has been argued that a sufficiently strong hybridization or Kondo interaction with respect to the
RKKY exchange interaction suppresses antiferromagnetic order to the benefit
of superconductivity. \cite{Monthoux2002,ThalmeierBook} 
General scaling behaviors of the characteristic energy scales
in Kondo lattice materials are a matter of intense debate,\cite{Nakatsuji2004, 
Curro2004, Yang2008, Yang2008a} and $T^*$ has traditionally been interpreted 
as the temperature below which the coherence of the Kondo lattice sets in. More
recently $T^*$ has been suggested to be the temperature scale denoting 
the development of a dense "Kondo liquid" within a "two-fluid" model also 
comprised of a "Kondo gas" phase of uncorrelated magnetic moments at 
high temperatures. This characteristic temperature scale is determined with many techniques 
such as thermodynamic, transport, knight shift measurements, and of course, quasielastic neutron 
 scattering \cite{Holland-Moritz1982, Murani1983, Severing1989, Lawrence2001, Murani1996}.
Another interesting scaling has been shown by Bauer \textsl{et al.} \cite{BauerPRL2004}: the 
superconducting transition temperatures $T_c$ of the CeMIn$_5$ and also of the PuMGa$_5$ family
vary linearly with the $c/a$ ratio of the tetragonal lattice constants, pointing towards the
importance of the anisotropic electronic structure for the 
superconducting state. This brings into focus the importance of the spatial distribution
of the crystal-field split Hund's rule ground state, which is highly 
anisotropic for materials containing rare earth.

The Hund's rule ground state of Ce$^{3+}$ with $J=5/2$ splits under the 
influence of a tetragonal crystal-field (point group $D_{4h}$)
into three Kramer's doublets, which can be represented in the basis 
of $|J_z \rangle$. The eigenfunctions of the three Kramer's doublets 
can be written as
\begin{eqnarray}
|2 \rangle &=\Gamma_6   =& \;\,\, |\pm1/2 \rangle\nonumber\\
|1 \rangle &=\Gamma_7^1 =& \beta |\pm 5/2\rangle - \alpha |\mp 3/2\rangle\label{EqCFScheme}\\
|0 \rangle &=\Gamma_7^2 =& \alpha |\pm 5/2\rangle + \beta |\mp 3/2 \rangle\nonumber
\end{eqnarray}
with $\alpha^2+\beta^2 = 1$. The anisotropy of certain wave functions may give 
rise to $4f$ conduction electron hybridizations which 
are more advantageous than others for forming a superconducting ground 
state.\cite{PagliusoPhysB320} The importance of momentum dependent 
({\bf q}-dependent) hybridization in these  
and some semiconducting Kondo materials has been investigated by several groups.
\cite{WeberPRB77,mena2005,GhaemiPRB75,BurchPRB75,KuboJPSJ75,MorenoPRL84} 

Various groups attempted to determine the crystal-field scheme of these compounds, but
there are significant discrepancies depending on the applied methods, which include
bulk measurements  based on transport, thermodynamic and NMR experiments.\cite{PagliusoPhysB320,CurroPRB64,TakeuchiJPSJ70,NakatsujiPRL89,ShishidoJPSJ71} 
Christianson et al. performed extensive inelastic neutron scattering
 (INS) studies,\cite{ChristiansonPRB70,ChristiansonPRB66} but phonon contributions in the energy range of the magnetic scattering,
 broadened crystal-field excitations due to hybridization effects and the enormous absorption 
of the sample's constituents make the determination of reliable magnetic intensities 
rather challenging. Since the latter give the wave functions via the 
transition matrix elements the resulting wave functions should be taken with care (see appendix) while
the transition energies are fairly sound (see section III B).   

We have shown for the case of the heavy fermion materials CePd$_2$Si$_2$ and CePt$_3$Si
that polarization dependent soft x-ray absorption (XAS) at the Ce M$_{4,5}$ edges 
can be complementary to neutron scattering in determining the 
ground state wave function.\cite{HansmannPRL100, WillersPRB80}  
XAS is highly sensitive to the initial state
and via its  polarization dependence (\textit{linear dichroism} $LD$)
direct information about the $|J_z \rangle$ admixtures of the ground state 
wave function can be obtained. Sensitivity to higher lying crystal-field states 
is achieved by thermally populating those states.\cite{HansmannPRL100, WillersPRB80}
 
We  present a combined inelastic neutron scattering 
 and soft x-ray absorption  study on the CeMIn$_5$ M = Co, Ir, and Rh compounds.
Combining both techniques has the advantage of determining transition energies 
and mixing parameters independently, each with the most suitable technique. 
The line positions in the magnetic contributions of the INS data yield 
the crystal-field transition energies within meV resolution, whereas the LD in XAS, 
when performed at temperatures where only the ground state is populated,
yields the ground state wave function, i.e. in case of a mixed ground state the 
mixing factor $\alpha^2$ (see eq. (1)) within $\Delta\alpha^2$ = $\pm 0.0025$. In the limit of 
small crystal-field splittings the dichroic signal of the ground state is independent 
of crystal-field energies. Once the ground state has been determined, the order 
of states can be determined from the temperature dependence of the LD since 
at finite temperatures it reflects the superposition of polarizations from 
each populated state, i.e. here the crystal-field energies, as determined from the 
neutron scattering experiment, enter via the thermal population. In addition,
high-resolution INS data are presented  to determine the characteristic
temperature T$^*$ for the \textsl{4f} conduction electron hybridization via the quasielastic 
line width. The latter has been applied successfully by several authors.\cite{Holland-Moritz1982, 
Severing1989, Lawrence2001, Murani1996} 

\section{Experimental and data correction}
The high quality single crystals of CeMIn$_5$ for the x-ray experiments 
were grown with the flux-growth method \cite{Monthoux2002}. The powder samples for the present 
INS experiments were the same samples as used by Christianson 
\textsl{et al.}\cite{ChristiansonPRB70,ChristiansonPRB66}.

\subsection{XAS}
The XAS spectra were recorded during various beam times at the two different 
synchrotron light sources  BESSY II and  NSRRC.
We recorded all  spectra with the total electron yield method (TEY) and under 
UHV, i.e. in a chamber with a pressure in the  10$^{-10}$ mbar range.  
Clean sample surfaces were obtained by cleaving the samples
{\it in situ}. At BESSY II we used the UE46 PGM-1 undulator beam line. 
The total electron yield (TEY) signal was normalized
to the incoming photon flux $I_0$ as measured at the refocusing
mirror.  The energy resolution at the cerium $M_{4,5}$ edges
($h\nu \approx 875-910$ eV) was set to 0.15 eV. The
undulator combined with a normal incident measurement geometry
allow for a change of polarization without changing the probed
spot on the sample surface. The two polarizations 
were $E\!\perp\!c$ and $E\!\parallel\!c$, $c$ being the long tetragonal axis.
At the NSRRC we performed the experiment  at the Dragon dipole beam line.  
The energy resolution at the cerium $M_{4,5}$ edges
 was set to  0.4 eV. The crystals were mounted with the $c$-axis perpendicular 
 to the Poynting vector of the light. By rotating the sample around this 
 Poynting vector, the polarization of the electric field can be varied 
 continuously from $E\!\perp\!c$ to $E\!\parallel\!c$. For all measurements 
 the sample was rotated 4 times by 90$^o$, so that for each orientation 
 $E\!\perp\!c$ and $E\!\parallel\!c$ two equivalent positions were measured.
Thus for both experimental end stations a reliable comparison of the 
spectral line shapes is guaranteed. We measured several crystals and/or recleaved 
in order to assure the reproducibility of the spectra (see table 1).

To calculate the XAS spectra we performed ionic full multiplet 
calculations using the XTLS 8.3 program\cite{TanakaJPSC63}. All 
atomic parameters are given by Hartree-Fock values,
with a reduction of about 40\%  for the $4f-4f$ Coulomb interactions
and about 20\% for the $3d-4f$ interactions to reproduce best the
experimental isotropic spectra, 
$I_{\rm isotropic} = 2 I_{\perp} + I_{\parallel}$. These values compare 
well with our findings for other heavy fermion compounds
\cite{HansmannPRL100, WillersPRB80} and account for the 
configuration interaction effects not included in the Hartree-Fock scheme.
Once the atomic parameters are fine tuned to the isotropic spectra, 
the polarized XAS data can be described by the incoherent sums of the 
respective polarization dependent spectra of the pure $|J_z \rangle$ 
states \cite{HansmannPRL100} as long as the crystal-field splitting $E_{CF}$
is small with respect to the spin orbit splitting $E_{SO}$. The latter requirement, 
which is fulfilled here ($E_{SO}$ $\approx$ 280~meV and $E_{CF}$ 
$\leq$ 30~meV), assures that interference terms resulting 
from intermixing of the $J=5/2$ and $J=7/2$ multiplet can 
be neglected. 
\begin{figure}[]
    \centering
    \includegraphics[width=1.0\columnwidth]{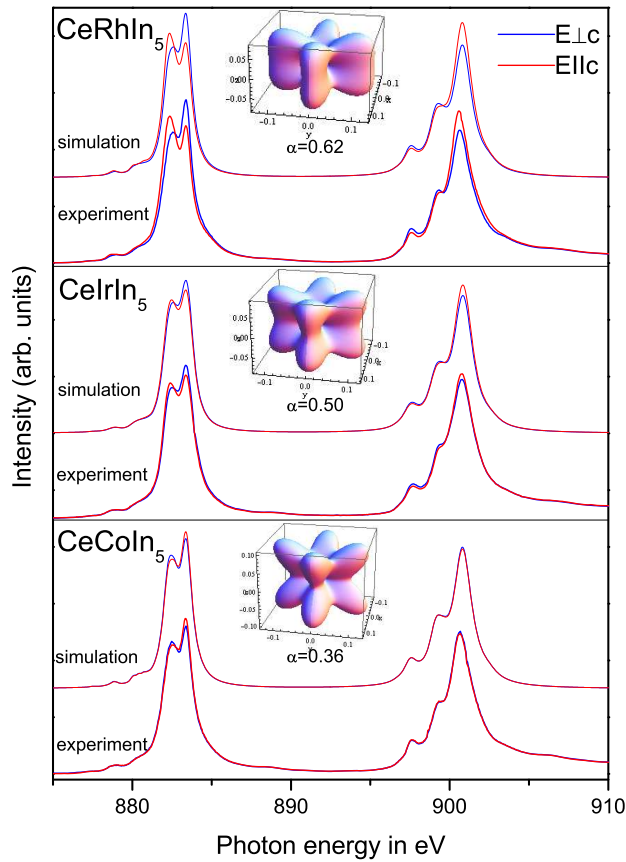}
    \caption{(color online) Low temperature (T = 20 K) linear polarized XAS spectra of 
    CeRhIn$_5$, CeIrIn$_5$, and CeCoIn$_5$ at 
    the Ce$^{3+}$ M$_{4,5}$ edges. The solid lines are the measured data, the dotted ones 
    the simulations as described in the text. The orbitals represent the spatial distribution of the 
    $4f$ wave functions according to the respective ground state admixtures, 
    $\alpha|\pm 5/2\rangle + \beta |\mp 3/2 \rangle$
    %\nonumber.
    }
    \label{LowT}
\end{figure}
\begin{table}
\begin{tabular}{llc}
\\ \hline
sample		 				& synchrotron facility			& sequence of temperatures [K] \\ \hline
CeCoIn$_5$ \#1		&NSRRC 	&78,12, 295 \\ 		
CeCoIn$_5$ \#2		&NSRRC 	&10 \\						
CeCoIn$_5$ \#3		&BESSY 	&18, 50, 80, 130, 17, 180, 280,  \\ & &17, recleave, 19, 50, 80, 280 \\	
CeIrIn$_5$ \#1		&NSRRC 	&18, 50, 80, 150, 300, \\ & &recleave, 300 \\ 
CeRhIn$_5$ \#2		&NSRRC 	&20, 80, 295 \\
CeRhIn$_5$ \#2		&BESSY 	&18, 50, 80, 150, 300 \\ \hline
\end{tabular}
\caption{Experimental details about the XAS measurements}
\end{table}

\subsection{Inelastic neutron scattering}

We have measured the neutron scattering function $S(Q,\omega)$ of 
CeRhIn$_5$, CeIrIn$_5$, and CeCoIn$_5$ with 
the inelastic time-of-flight spectrometer HET at the neutron spallation source 
ISIS with incoming energies of 20 and 60 meV and energy resolutions of 0.6 and 1.8 meV 
in the 2.5 m detector banks.  The low angle banks cover 
2$\theta$= 9$^{\circ}$ to $29^{\circ}$ and the high angle banks 
130$^{\circ}$ to 140$^{\circ}$. All low  and high angle detectors 
are grouped together respectively in order to gain statistics. We therefore refer to 
S(2$\theta,\omega$) from now on. The corresponding averaged momentum 
transfers at elastic position for the low and high angle groupings are 
$\overline{Q}$ $\approx 1.8 \AA$ and $\approx 10.0 \AA$ 
for the 60 meV and $\overline{Q}$ $\approx 1.0 \AA$ and $\approx 5.5 \AA$ for the 20 meV data. 
Some data were taken at the cold time-of-flight spectrometer IN6 at ILL 
with an incoming energy of 3.1 meV and an energy resolution of 70 $\mu$eV at 
elastic position. All detectors from 10$^{\circ}$ to 115$^{\circ}$ have been grouped 
together. Because of the small incident energy the momentum transfer Q is $\leq$2 \AA ~in the energy 
window of interest. A flat sample geometry was used for all neutron experiments 
and -- because of the enormous absorption of In, Rh, and Ir -- well defined, but thin samples 
were crucial in order to guarantee transmissions of at least 30\%. This reduced the sample
amount for the IN6 experiment to about 5 g.  All data have been normalized to monitor 
count rate and vanadium and have been corrected for absorption and self-shielding. 
The description of the phonon correction has been moved to the appendix.

\section{Results}
For Ce$^{3+}$ in D$_{4h}$ point symmetry the crystal-field Hamiltonian 
$H_{CF}=B_2^0 O_2^0 + B_4^0 O_4^0 + B_4^4 O_4^4$ describes the 
crystal field potential when the three Stevens parameters $B_2^0$,
$B_4^0$, and $B_4^4$ are determined. The $B_k^m$ parameters are determined via the 
crystal-field transition energies within the Hund's rule ground state 
and the mixing parameter $\alpha$. In section \textbf{A} the mixing parameter will 
be determined from the LD in the XAS data at low temperature, in
section \textbf{B} the crystal-field energies are obtained from thermal 
neutron scattering data, and \textbf{C} the order of crystal-field states 
will be confirmed from the temperature of the LD effect in the XAS spectra. 
In section \textbf{D} we finally present the quasielastic results 
from the cold neutron data in order to determine the hybridization temperatures T$^*$. 
\begin{figure*} 
    \centering
    \includegraphics[width=2.0\columnwidth]{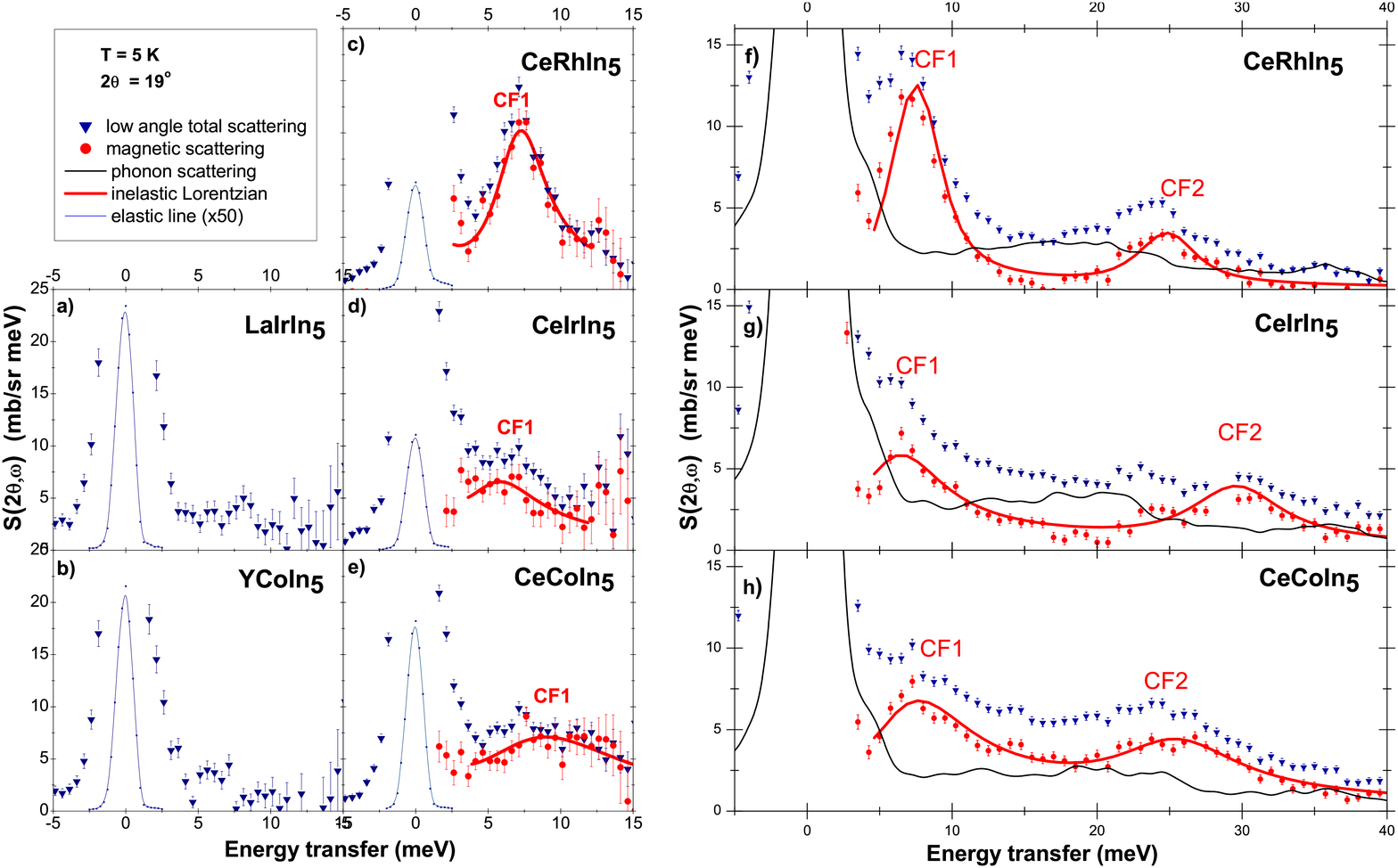}
    \caption{(color online) Inelastic neutron data at 5 K with 20 meV a)-e) and 
    60 meV f)-h) incident energy. The blue triangles represent the total scattering, the 
    black lines the phonon intensities as obtained from a high to low 
    angle scaling. The red circles show the magnetic scattering obtained 
    after phonon correction (see appendix) and the red lines are Lorentzian fits.} 
    \label{HETData}
\end{figure*}   
\begin{table*}[ht]
\begin{center}
\begin{tabular}{l@{\hskip1.0cm}c@{\hskip1.0cm}c@{\hskip1.0cm}c@{\hskip1.0cm}c@{\hskip1.0cm}c@{\hskip0.8cm}}
\\ \hline
[meV]& CeRhIn$_5$   & CeIrIn$_5$   & CeCoIn$_5$& instrument & CeAu$_2$Si$_2$\cite{Severing1989, Severing1989a}  \\ \hline

$\Gamma_{qu}$/2$@$ 8K  &   1.2 $\pm$ 0.2&            &           &     IN6 3.1 meV& 0.13  \\
$\Gamma_{qu}$/2$@$ 75K &   1.7 $\pm$ 0.2&  2.7 $\pm$ 0.5 &  3.9 $\pm$ 0.5&     IN6 3.1 meV& 0.43 \\ \hline

E$_1$                &   7.0 $\pm$ 0.5& 5.2 $\pm$ 1.0  & 6.8 $\pm$ 1.0 &     HET 20meV  & 16.5  \\
E$_2$                &  24.7 $\pm$ 1.0&29.4 $\pm$ 1.5  &25.0 $\pm$ 1.5 &     HET 60meV  & 21.0 \\

$\Gamma_{in}^1$/2$@$ 5K &   1.6 $\pm$ 0.5&  3.0 $\pm$ 0.5 & 4.8 $\pm$ 0.8&    HET 20/60 meV                 \\
$\Gamma_{in}^2$/2$@$ 5K &   1.8 $\pm$ 0.5&  3.4 $\pm$ 0.8 & 4.7 $\pm$ 0.8&    HET 60 meV                     \\ 
$\Gamma_{in}^1$/2$@$75K&    2.5 $\pm$ 0.5&  3.7 $\pm$ 0.5 &  4.4 $\pm$ 0.8&   IN6 3.1 meV                    \\ \hline
 
B$_{20}$             &  -0.928    &  -1.197    & -0.856    &                                       \\
B$_{40}$             &   0.052    &   0.069    &  0.063    &                                          \\
$|B _{44}|$          &   0.128    &   0.088    &  0.089    &                                      \\ \hline
\end{tabular}
\caption{crystal-field energies E$_{1,2}$, widths (HWHM) $\Gamma$/2, and crystal-field parameters are given in meV.
The fifth column gives instrument and incident energy E$_{in}$ from which the parameters were determined. 
Values of CeAu$_2$Si$_2$ are shown for comparison. The crystal-field parameters from the full 
multiplet calculation are given in Stevens formalism.}
\end{center}
\end{table*}

\subsection{Low temperature polarized soft XAS: the ground state wave function}

Figure \ref{LowT}  shows the low-temperature 20 K linear polarized XAS data 
of CeRhIn$_5$, CeIrIn$_5$, CeCoIn$_5$ at the Ce$^{3+}$ $M_{4,5}$ 
edge full lines), i.e. at a temperature sufficiently low so that 
the ground state is populated. The latter has been verified with
the knowledge of the crystal-field energies (see table 2). For CeCoIn$_5$ the 
linear dichroism (LD) is smallest, but has an opposite sign with 
respect to  CeIrIn$_5$ and CeRhIn$_5$. CeRhIn$_5$ has the largest 
LD. Figure \ref{LowT} shows further the simulations based on a full multiplet treatment
as described in the experimental section (dotted lines).
 For CeCoIn$_5$ we find from our full multiplet calculation
$\alpha=0.36 $. This value is just below the 
zero polarization for  $\alpha=\sqrt{\frac{1}{6}}\approx 0.41$ where the LD vanishes 
when it changes sign. For CeIrIn$_5$ and CeRhIn$_5$ we obtain 
$\alpha=0.50 $ and $\alpha=0.62$, respectively. The resulting 
ground state wave functions, with an arbitrarily chosen phase, are:
\begin{eqnarray*}
\mbox{CeRhIn$_5$:} \quad |0 \rangle =\Gamma_7^2 = 0.62 |\pm5/2\rangle + 0.78 |\mp3/2\rangle\\
\mbox{CeIrIn$_5$:} \quad |0 \rangle =\Gamma_7^2 = 0.50 |\pm5/2\rangle + 0.87 |\mp3/2\rangle\\
\mbox{CeCoIn$_5$:} \quad |0 \rangle =\Gamma_7^2 = 0.36 |\pm5/2\rangle + 0.93 |\mp3/2\rangle
\end{eqnarray*}%
The orbitals shown in Figure \ref{LowT} show the spatial distributions of the 
$4f$ electrons for the respective crystal-field ground
states. The higher the $|5/2\rangle$ contribution to the ground state the flatter 
the $4f$ distribution, i.e. CeRhIn$_5$, which does not become superconducting 
at ambient pressure, has the flattest $4f$ orbital. The pure 
$|5/2\rangle$ orbital is donut and the pure $|3/2\rangle$ is yo-yo shaped 
(see e.g. Willers \textsl{et al.}\cite{WillersPRB80}). 
The general trend of a decreasing $|5/2\rangle$ contribution to the ground state 
from M = Rh, Ir to Co agrees with the INS findings by Refs. \onlinecite{ChristiansonPRB70,ChristiansonPRB66} 
but we observe smaller $|5/2\rangle$ contributions (note: $\sqrt(1 - \alpha^2)$ 
= $\beta$ as given by Christianson \textsl{et al.}). This is most likely due to systematic
errors in the phonon correction of the INS neutron data (see appendix). 

\subsection{INS: crystal-field transition energies}
 
Figure~\ref{HETData} shows the scattering function $S(2\theta,\omega)$ at T = 5 K for
small scattering angles 2$\overline{\theta}$= 19$^{\circ}$ measured with two 
incident energies, $E_{in}$ = 20 meV a) to e) and 60~meV f) to h). The 
separation of magnetic and phonon correction has been performed as 
described in the appendix. $E_{in}$ = 20 meV: a) and b) exhibit the 
scattering function $S(2\theta,\omega)$ of two non-magnetic reference samples 
and c) to d) show $S(2\theta,\omega)$ of the three cerium compounds. 
The blue triangles are the total scattering. The data of the La samples show that 
there is only little phonon scattering for 20 meV incident energy. 
It has nevertheless been corrected for (method 2) in order to determine the magnetic 
scattering in the cerium data. The red circles are 
the pure magnetic scattering after phonon correction. $E_{in}$ = 60 meV: the 
blue triangles in f) to h) are the total scattering for small scattering angles, 
i.e. for low $Q$ and the black lines reflect the phonon 
scattering as obtained from the high to low angle scaling method. 
The phonons scale with R = 1/8 from 2$\overline{\theta}$= 135$^{\circ}$ to 
2$\overline{\theta}$= 19$^{\circ}$ as empirically found 
from the non-magnetic reference samples (see appendix).
The red circles are the pure magnetic scattering in the low angle banks, 
resulting from subtracting the scaled phonon intensities (black lines)
 from the total scattering (blue triangles). 
 
 All three compounds exhibit two magnetic 
 ground state excitations at about 5-7 and 24-30~meV. 
The low energy one (CF1) is best resolved in the 20 meV data while the 60 meV data 
show both excitations (CF1 and CF2). The quasielastic scattering cannot be resolved 
with these thermal measurements. The inelastic, magnetic excitations have been 
described with two inelastic Lorentzians. The crystal-field transitions 
are fairly sharp in CeRhIn$_5$ but much broader for CeIrIn$_5$ 
and CeCoIn$_5$, leading to considerable error bars in the line widths and position. 
The 60~meV data are described with two inelastic Lorentzians of same widths, but only the parameters 
of the higher energy excitations were freely varied, position and width of the 
low energy excitations have been determined from the 20~meV data. We can state that 
the excitations broaden from M = Rh to Ir and Co, consistent with the expected 
increase in the electronic specific heat coefficient $\gamma \approx 1/T_K$ from 
Rh to Ir to Co. Table 2 gives the summary of all parameters.

\begin{figure*}[]
    \centering
    \includegraphics[width=2.0\columnwidth]{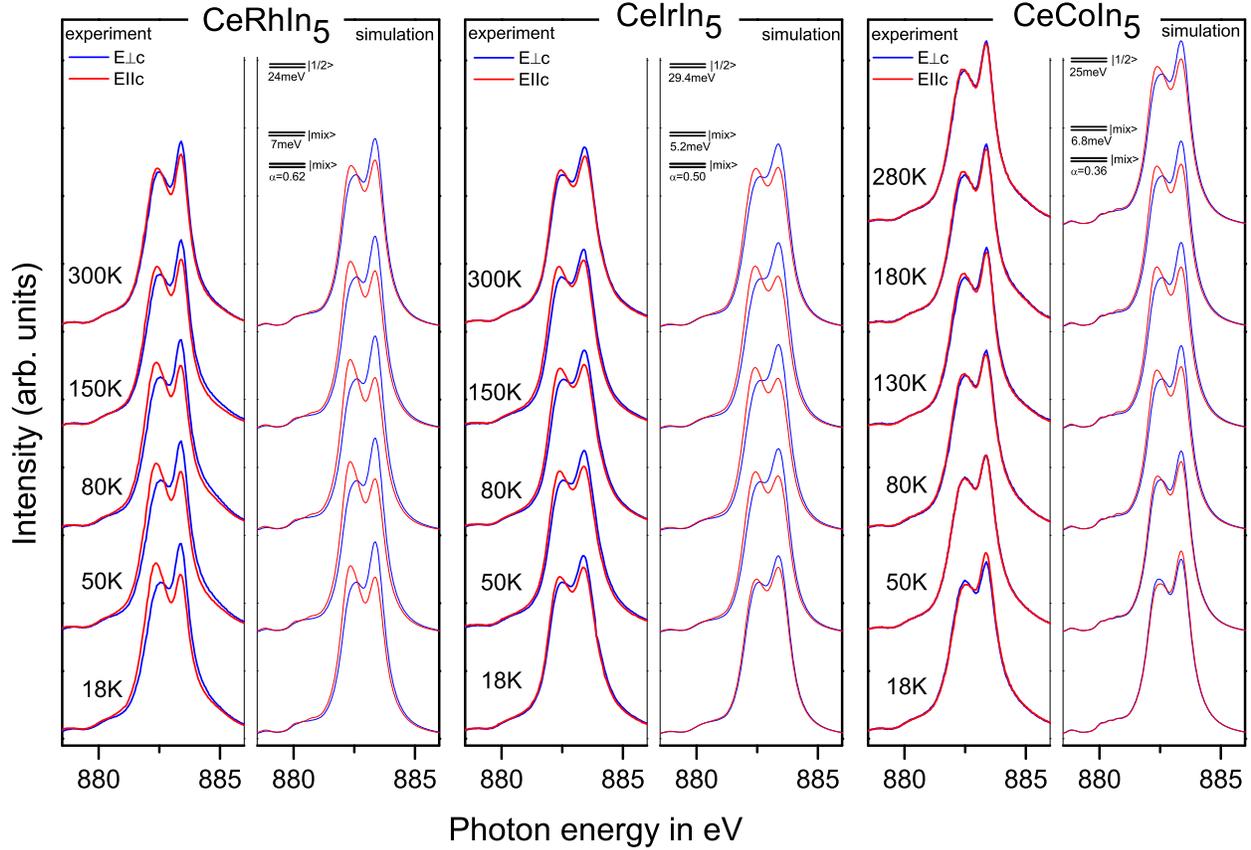}
    \caption{(color online) Left panel: Temperature-dependent linear polarized XAS spectra
    at the Ce$^{3+}$ $M_5$ edge of CeRhIn$_5$, CeIrIn$_5$, and CeCoIn$_5$. 
    The solid lines correspond to the measured data, the dotted lines to the
    simulation based on the crystal-field energies from the neutron data and the ground state
    wave functions from the low temperature XAS data.}
    \label{AllT}
\end{figure*}

In the $E_{in}$ = 60 meV data phonon and magnetic scattering are very much intermixed, 
the line widths are broad and we know from the non-magnetic reference samples 
that there is a substantial error in not 
taking out enough phonon scattering below 10 meV, i.e. in the range of the lower crystal-field 
excitation (this is explained in the appendix, see figure \ref{Y-La_60meV}). We know 
further that the fine structure of the phonon scattering is not ideally represented by the $Q$ scaled high angle data 
(see appendix, right hand panel of figure \ref{Y-La_60meV}). Hence we do not attempt to determine magnetic
intensities to obtain transition matrix elements. 
The magnetic signal is nevertheless strong enough to
determine the crystal-field energies fairly well, irrespective of the method of phonon correction as can be
seen from the fact that we obtain similar line positions as Christianson et al. who applied some extra 
scaling factor for the phonon correction in addition to the one found from the 
La samples (compare appendix with Refs.  \onlinecite{ChristiansonPRB70,ChristiansonPRB66}). 
However, for CeIrIn$_5$ Christianson et al. can only state that the low lying crystal-field state must be 
some where between 0 and 7 meV. Our neutron data show, due to the better flux and resolution, 
that the level is located between 5.2 $\pm$ 1 meV (see figure~\ref{HETData} e)). This is supported by the fact that the 
polarization effect of the XAS data increases from 20 to 80 K (see section C, figure~\ref{AllT}), i.e.
when the thermal occupation of the first excited state takes place, before it 
decreases again due to the beginning population of the second excited state. 

We find narrower line widths and we find that the width of the spectra increases
in the sequence M=Rh to Ir and becomes broadest for Co in 
contrast to Refs. \onlinecite{ChristiansonPRB70,ChristiansonPRB66}. We believe that this 
is due to the better signal to noise ratio of the present data. While we do 
not consider the quality of the magnetic intensities to be sufficiently accurate 
to determine the wave functions properly from these data, we agree with Christianson $et al.$
\onlinecite{ChristiansonPRB70,ChristiansonPRB66} that general intensity considerations 
based on the selection rule $\Delta J_z$ = $\pm$1 lead to the conclusion that 
the $|5/2\rangle$ contribution to the ground state decreases from Rh to Ir and 
Co, that is in agreement with our low temperature XAS data. We
conclude further from the intensity ratios, as Chrstianson $et al.$
\onlinecite{ChristiansonPRB70,ChristiansonPRB66}, that the second 
excited state is the pure $|1/2 \rangle$ in all compounds.

The same set of data was taken at T = 120~K (not shown here) where we would expect to see the 
transition from the first $|1 \rangle$ to the second $|2 \rangle$ excited state 
at $E_2$-$E_1$, but we do not resolve another peak. However, it 
should be noted that at 120~K the phonon correction is more important with respect 
to 5~K while the magnetic lines are broader. In addition, this third transition is weaker than
the two ground state excitations.   
\begin{figure}[]
    \centering
    \includegraphics[width=1.0\columnwidth]{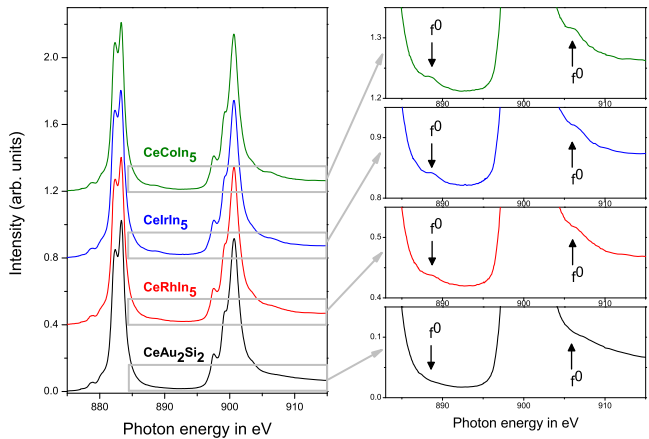}  
    \caption{(color online) Isotropic spectra of CeMIn$_5$ with M=Rh, Ir and Co, (at 20 K)
    and of CeAu$_2$Si$_2$ (at 40 K) for comparison. All spectra were recorded under 
    the same beam line conditions at the NSRRC. In the right panel are blow-up graphs 
    with arrows pointing out the $4f^0$ initial state  contributions.}
    \label{isotrope}
\end{figure}
\subsection{Temperature dependent polarized soft XAS: sequence of crystal-field states}

When analyzing the temperature dependence of the polarization dependent XAS data we will only show
 the $M_5$ edge for clarity. Figure \ref{AllT}  
shows the temperature dependence of the LD for all three cerium compounds. For each compound
the measured data and corresponding simulations are shown. 
For CeRhIn$_5$ the LD in the experimental data increases slightly from 18 to 80 K 
indicating the population of the first excited
crystal-field level and then decreases with further rising temperature due to 
population of the next state. When all states are equally populated the polarization should 
vanish entirely since a equal occupation resembles the fully degenerate Hund's rule ground state which has 
spherical symmetry. To the right of the measured data the 
simulated XAS spectra are shown. There, the crystal-field energies as determined in 
the neutron experiment and the mixing factor $\alpha$ as obtained from the 
low temperature XAS data are used as input parameters. The $|1/2 \rangle$ has been assumed to be
the highest lying crystal-field level and only the population of states 
has been adapted to the corresponding temperature. The simulation reproduces 
well the trend of the temperature dependence of the LD, i.e. it increases at first and then decreases
above 80 K with further rising temperature. The assumption of a different order of states would lead
to a change of sign in the polarization for increasing $T$, which is in 
contradiction to the observation and can therefore be excluded. 
%In \cite{HansmannPRL100, WillersPRB80} the polarizations of the pure $|5/2 \rangle$ states are shown
The other panels of figure \ref{AllT} show the same set of spectra 
for CeIrIn$_5$ and CeCoIn$_5$. For CeIrIn$_5$ we observe very much the same effect as for CeRhIn$_5$ and we are able to 
simulate the general trend of the temperature dependence in the same manner. For CeCoIn$_5$ 
the temperature dependence seems different, but is
based on the same effect, namely occupation of the next higher states: the low temperature 
polarization has a different sign with respect to M= Rh and Ir (see section \textbf{A}) 
so that here the occupation of the first excited state leads to a decrease of polarization 
with rising temperature and a change of sign between 50 and 80 K. Then the LD 
increases at first with further rising temperature and starts to decreases 
again above 180 K. Here too we are able to simulate this temperature trend under the assumption
that the $|1/2 \rangle$ is the highest lying crystal-field state. For M=Co, 
the assumption of a different order of states would not give rise to a change 
of sign in the polarization as function of temperature. 
This analysis of the temperature dependence is analogous to our previous 
results on CePt$_3$Si.\cite{WillersPRB80} Ground state, energy splittings, 
and order of states describe the crystal-field potential fully and the corresponding 
crystal-field parameters (in Stevens formalism) are listed in table 2.  

While the temperature dependence of the LD is qualitatively 
reproduced with our simulations, there is also some quantitative deviation:
 at high temperatures the measured LD is always smaller 
than the simulated one. We exclude depolarization effects due to a) surface
degrading and/or b) polycrystalline contributions. Possibility a) can be excluded since 
we recleaved and repeated the measurements to assure reproducibility (see table 1) and we exclude b)
since it would require an unrealistic 35\% of polycrystalline 
contribution in order to account for the mismatch in e.g. the Co data at 280 K. 
More physical and interesting is to consider Kondo interactions: 
they will have a depolarizing effect too. The 
hybridization of the $4f^1$ state with the surrounding conduction 
band has not been considered in the present analysis of the XAS data. 
Yet, the existence of the latter can be seen from the $4f^0$ initial state satellites at the 
high energy tails of the M$_{4,5}$ edges. Figure ~\ref{isotrope} shows 
the isotropic spectra and the tails of the M$_{4,5}$ edges on a blown-up scale 
for the CeMIn$_5$ compounds and in comparison for CeAu$_2$Si$_2$. 
The arrows in Figure ~\ref{isotrope} indicate the position of the 
$f^0$ spectral weight. CeAu$_2$Si$_2$ is an antiferromagnet with a 
small hybridization temperature ($T^* = 1.5$ K)\cite{Severing1989} 
and with a good agreement of simulated and measured XAS data at all 
temperatures \cite{mythesis}. It is interesting to note that the $4f^0$ 
spectral weight is basically non-existent in CeAu$_2$Si$_2$, but stronger 
in the CeRhIn$_5$ data and again more pronounced for CeIrIn$_5$ and CeCoIn$_5$. 
In the same sequence the inelastic line widths in the INS spectra and 
the deviation between simulated and measured spectra increase. We
speculate that hybridization effects, which are not yet included in the 
calculation, may be responsible for these quantitative discrepancies. 
It is desirable that further theoretical work be carried out, using for 
instance the Anderson impurity model, to find out how much the CF wave 
functions are modifed from our present estimates. We nevertheless expect 
that these modifications are very modest for the CeRhIn5 since the $f^0$ 
weight in the ground state is minimal. We may even speculate that the 
corrections are also small for the CeCoIn$_5$ and CeIrIn$_5$ in view 
of the special condition that the crystal-field ground state wave 
function is very close to cubic, i.e. almost isotropic.

\subsection{INS: hybridization temperature T$^*$}

\begin{figure*}
    \centering
    \includegraphics[width=2.0\columnwidth]{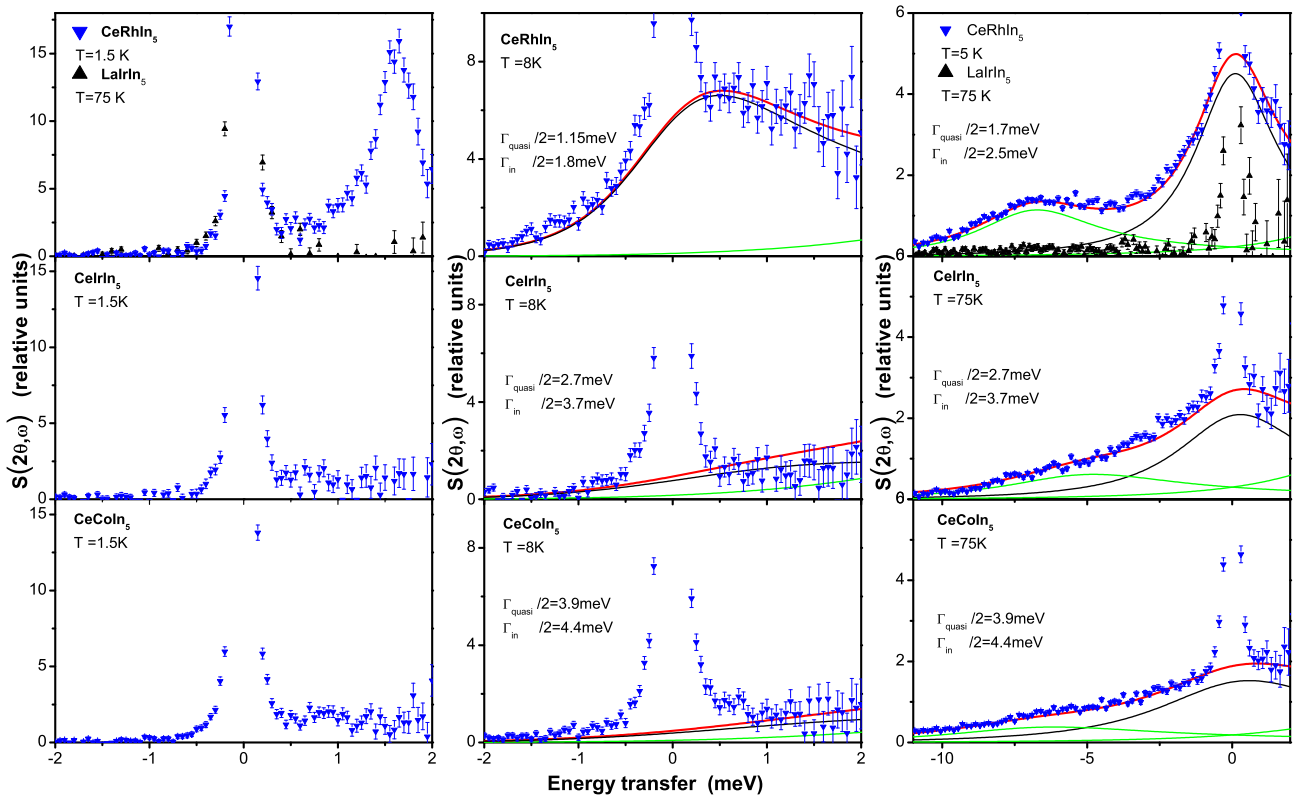}
    \caption{(color online) Neutron data with 3.1 meV incident energy. For T = 1.5 and 8 K 
    the neutron energy loss, for 75 K the neutron energy gain side is shown. The blue triangles 
    (pointing down) are the total scattering of the cerium samples, the black triangles 
    (pointing up) are the La scattering. The 75 and 8 K data are described with one 
    quasi- (black) and one inelastic Lorentzian (green) with intensity ratios according to 
    the crystal-field models (see table 2). The red lines reflect the total, fitted 
    magnetic scattering. The line widths are discussed in the text.}
    \label{IN6}
\end{figure*}

Figure~\ref{IN6} shows high resolution inelastic neutron scattering data of 
CeMIn$_5$ M=Rh, Ir and Co, taken with IN6 at ILL with E$_{in}$ = 3.1 meV 
incident energy at 1.5, 8, and 75 K. The blue triangles (pointing down) are 
the total scattering from the cerium samples. We further show 
the scattering from LaIrIn$_5$ at 75 K (see black triangles pointing up in the 1.5 and 
75 K CeRhIn$_5$  spectra) in order to verify that the phonon scattering is negligible 
at these temperatures in this energy window. At 1.5 K CeRhIn$_5$ is magnetically ordered (T$_N$=3.8~K)
and the scattering function exhibits in addition to the elastic line 
(incoherent, elastic, nuclear scattering) some magnon density of states 
which peaks at about 1.7 meV. At 8~K the spectrum consists mainly of 
quasielastic scattering, which is well described with a quasielastic Lorentzian 
with HWHM $\Gamma$/2 = 1.2~($\pm$0.1)meV. Note, that for $\hbar\Gamma/2 < k_BT$ 
a quasielastic Lorentzian appears highly asymmetric in energy \cite{Fulde1986}. 
The crystal-field excitation CF1 contributes only little in this energy window but 
has nevertheless been taken into account according to the crystal-field model (see green line). 
A quasielastic line width of $\Gamma$/2 = 1.2~($\pm$0.1)~meV corresponds with $\hbar\omega$ = k$_B$T to
T$^*$~$\approx$~14~($\pm$1)~K. At 1.5 and 8 K the scattering intensity for CeCoIn$_5$ and CeIrIn$_5$ is 
considerably lower with respect to CeRhIn$_5$, but not zero,  so that we conclude the 
quasielastic lines are broader and peaked outside the energy window available at low temperatures. 
It should be mentioned that although CeCoIn$_5$ is in the superconducting phase at 1.5 K, we did not observe 
 the spin resonance at 0.6~meV \cite{Stock2008}, most likely because 
 this resonance is fairly sharp in reciprocal space whereas the present data are averaged over all $Q$.  

Before discussing the CeCoIn$_5$ and CeIrIn$_5$ data at 1.5 and 8 K data in more detail,  
we look at the high resolution spectra at 75 K (right hand panel of figure~\ref{IN6}). 
Here the energy window is larger due to population of states on the neutron energy gain side. 
Thanks to population and resolution the quasielastic scattering and the low lying 
crystal-field excitation can be observed simultaneously in the same energy window. The black triangles 
in the CeRhIn$_5$ spectrum are the scattering from LaIrIn$_5$ at 75 K. Since it 
is almost negligible we will not consider it when describing the cerium data. For 
CeRhIn$_5$ quasi- and inelastic scattering are well resolved while for CeIrIn$_5$ and
CeCoIn$_5$ the spectra are broader so that it is difficult to separate the two. 
We fit the data with the crystal field model, i.e. keep the intensity ratios 
of quasi- and inelastic lines as well as the line positions fixed (see parameters in table 2). 
Thus only the line widths and an overall intensity parameter were varied. The latter is necessary although
we scaled to absolute intensities, most likely since the absorption corrections are so large. %This means
%that any hole or larger grain in the powder sample falsifies the correction and hence the intensity. 
The data are well described with our crystal-field parameters, the resulting quasi- and inelastic 
line widths at 75 K are listed in table 2. 

We return to the 8 K data in order to give an estimate for the quasielastic
line widths of CeCoIn$_5$ and CeIrIn$_5$. The quasielastic intensity at 8 K is 
scattering from the crystal-field ground state which is proportional 
to the square of the transition matrix element. The ground state of each compound is know from 
the low temperature XAS data, hence we know the respective matrix elements of the 
quasielastic scattering. We further know the relative scaling factor of the data from 
the 75 K spectra. For CeRhIn$_5$ the quasileastic width and intensity can easily be fitted. Knowing
scaling factors and matrix elements allows us then to estimate
the line widths of the Co and Ir samples at 8 K (middle columns of figure~\ref{IN6}).
 We use the widths at 75 K as a crude guess for the quasielastic line widths of 
 CeCoIn$_5$ and CeIrIn$_5$ at 8 K and take into account the inelastic scattering 
 according to the crystal-field model. The red lines are the result. The agreement is 
reasonable so that we give the following values for the corresponding coherence 
temperatures T$^*$: 
\begin{eqnarray*}
\mbox{CeRhIn$_5$:} \quad T^* = 14  \pm 1 K\\ 
\mbox{CeIrIn$_5$:} \quad T^* \approx 30 \pm 5 K\\
\mbox{CeCoIn$_5$:} \quad T^* \approx 45 \pm 8 K
\end{eqnarray*}

\section {Discussion}

The hybridization temperatures T$^*$ as determined from the present neutron scattering experiments
are smallest for CeRhIn$_5$ and largest for CeCoIn$_5$ in agreement with the increasing spectral
weight of the $f^0$ contribution to the XAS spectra. The values we find from neutron scattering
agree well with temperatures below which Fermi liquid behaviour sets in: the anomalous Hall effect  below
20 K for CeRhIn$_5$, 31 K for CeIrIn$_5$, 53 K for CeCoIn$_5$ \cite{Yang2008}. Knight shift experiments give 
10-12 K for CeRhIn$_5$, and for CeCoIn$_5$ 42 K for in plane and 89-95 K for out of plane, 
i.e. a powder averaged value of about 58 K \cite{Curro2004}.
Thermal and transport measurements by Nakatsuji \textsl{et al.} give 
T$^*$ $\approx$ 45 K for CeCoIn$_5$ and the entropy of the specific heat of CeRhIn$_5$ 
reaches $1/2R \ln{2}$ at about 10-12 K \cite{HeggerSCinCeRhIn5}. Our findings are further 
in agreement with the findings of the Fermi surfaces. While the general features of the 
Fermi surface of CeRhIn$_5$ are more like the Fermi surface of LaRhIn$_5$, which 
has no $4f$ electrons, the Fermi surfaces of CeCoIn$_5$ and CeIrIn$_5$ are well 
described with a more itinerant $4f$ band model \cite{ShishidoJPSJ71}. A summary 
of values and references can be found in the supplementary information of Ref.  \onlinecite{Yang2008a}. 

CeRhIn$_5$ is the most localized member of this family. The hybridization temperature of CeRhIn$_5$ compares
with those of other heavy fermion materials like CeCu$_2$Si$_2$ (T$^*$ $\approx$ 10 K) \cite{horn1981}, 
CeRu$_2$Si$_2$ (T$^*$ $\approx$ 10 K) \cite{Severing1989}, and with the ones of the antiferromagnetic compounds
CePd$_2$Si$_2$ (T$^*$ = 10 K, T$_N$ = 8 K)\cite{Severing1989} 
and the cubic parent compound CeIn$_3$ (T$^*$ $\approx$ 10 K, T$_N$ = 10 K)\cite{Knafo2003}. With the 
exception of CeRu$_2$Si$_2$ all these compounds, exhibit superconductivity: CeCu$_2$Si$_2$ at ambient pressure 
depending on sample stoichiometry, or CeRhIn$_5$, CePd$_2$Si$_2$ and CeIn$_3$ with an applied pressure of 
 1.6 GPa and about 2.5 GPa for the latter two\cite{HeggerSCinCeRhIn5, Grosche1996, MathurNature394}. It is intriguing that for 
CeAu$_2$Si$_2$, which also orders antiferromagnetically at T$_N$ $\approx$ 10 K but has a much 
smaller hybridization temperature of T$^*$ = 1.5 K\cite{Severing1989}, no superconductivity has been reported 
up to 17 GPa \cite{link1997}. Since pressure on the cerium ion increases delocalisation, these findings 
underline the idea that sufficient Kondo screening favors superconductivity 
to the detriment of antiferromagnetic order \cite{Monthoux2002,ThalmeierBook}. 
For CeIrIn$_5$ and CeCoIn$_5$ where the hybridization
temperatures are larger the Kondo screening seems to be sufficiently large so that superconductivity can develop 
at ambient pressure. Here the larger hybridization temperature
of CeCoIn$_5$ (T$_c$ = 2.3 K) with respect and CeIrIn$_5$ (T$_c$ = 0.4 K) goes along with a higher superconducting 
transition temperature. 

There is another aspect which makes CeIrIn$_5$ and CeCoIn$_5$ remarkable. 
They are rare examples for compounds where the hybridization temperature T$^*$ 
is similar in magntitude to the size of the (low lying) crystal-field splitting. 
One may speculate that this should have an effect on the degeneracies 
involved when describing ground state properties.

Along with the increasing hybridization from M=Rh, via Ir 
to Co goes a decrease of the $|5/2\rangle$ contribution to the ground state.
CeRhIn$_5$ which has the flattest $4f$ orbital (see orbitals in figure 2) does not 
become superconducting at ambient pressure and it does not appear in the T$_c$ 
versus $c/a$ scaling plot as suggested by Pagliuso\cite{PagliusoPhysB320}, 
although its $c/a$ ratio is in between the values of CeIrIn$_5$ and 
CeCoIn$_5$. The CeIr$_{1-x}$Rh$_x$In$_5$ and CeCo$_{1-x}$Rh$_x$In$_5$ systems, 
however, exhibit superconductivity and fit into this $c/a$ scaling. It would 
be interesting to see where, in a similar scaling of T$_c$ with the
 $4f$ wave functions, the CeIr$_{1-x}$Rh$_x$In$_5$ 
 and CeCo$_{1-x}$Rh$_x$In$_5$ would fit. Since the crystal-field energies do not
 vary much from sample to sample it would be sufficient to determine the 
 ground state wave functions with linear polarized XAS
 at low temperatures.

\section{Summary}
We have determined the hybridization temperatures T$^*$ and 
crystal-field schemes of CeMIn$_5$ M=Rh, Ir and Co with inelastic neutron 
scattering and polarized soft x-ray absorption. The hybridization temperatures T$^*$ as
determined from the line widths of the inelastic neutron data increase from M = Rh to Ir, 
and are largest for Co which supports the idea that increasing Kondo interaction favours 
superconductivity while preventing long range magnetic order. The 
hybridization temperature of CeRhIn$_5$, the most localized member 
of the family, is comparable to the CeRu$_2$Si$_2$ and CeCu$_2$Si$_2$. 
For CeIrIn$_5$ and CeCoIn$_5$ the energy scale of the $4f$ conduction electron interaction is of the order 
of the energy of the low lying crystal-field excitation, which may have
an impact on the ground state degeneracy and/or properties. Our finding of the crystal-field schemes 
is coherent with previous work by Christianson \textsl{et al.} \cite{ChristiansonPRB70,ChristiansonPRB66} 
but we can give more precise values for the ground state wave functions from our XAS data. 
We find that the $|5/2 \rangle$ contribution to the ground state is largest(smallest)
for CeRhIn$_5$ (CeCoIn$_5$) so that CeRhIn$_5$ has the flattest $4f$ orbital. 

\begin{figure}[]
    \centering
    \includegraphics[width=1.0\columnwidth]{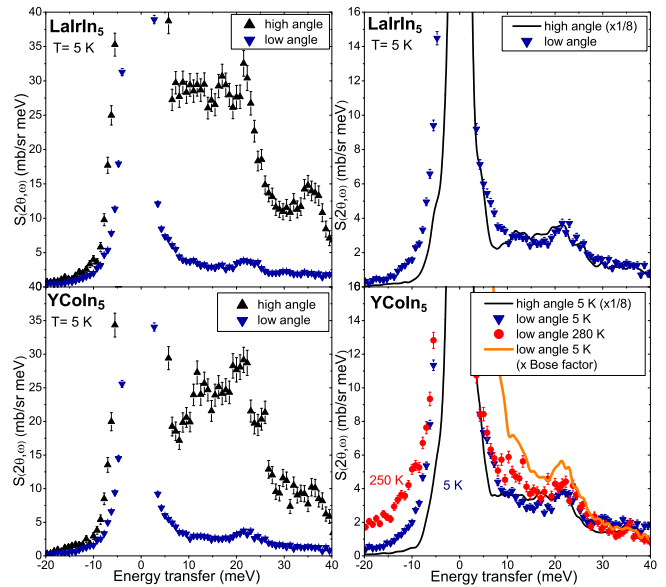} 
    \caption{(color online) Inelastic neutron scattering data of the non-magnetic 
    reference compounds YCoIn$_5$ and LaIrIn$_5$. \textsl{Left}: high 
    and low angle data at 5 K. \textsl{Right}: 5 K low angle data shown on an 
    expanded intensity scale and scaled ($R=1/8$) 
    high angle data. \textsl{Bottom right}: the red squares are the same low angle data taken at 280 K. 
    The orange line represents the 5 K data scaled to 250 K with the Bose factor.}
    \label{Y-La_60meV}
\end{figure}

\section{Appendix}
A single phonon scattering process increases with momentum transfer $Q$ as $Q^2$ and
magnetic scattering decreases with the magnetic form factor of the magnetic ion. Phonon
scattering follows Bose and the occupation of crystal-field states is 
according to Boltzmann statistics. We shall now discuss the two ways of separating magnetic and
phonon scattering: 1) subtraction of the high~$Q$ data from the low~$Q$ data after 
scaling the high~$Q$ data with a scaling factor $R$ which 
has been determined with a non-magnetic reference compound\cite{Murani1983} and 
2) subtraction of $S(Q,\omega)$ of a non-magnetic reference which 
has been scaled by the average scattering cross-section. Method 1): since we group angles 
and not momentum transfers $Q$ we will determine the scaling factor for high to low 
scattering angles $2\theta$ and not for $Q$. The left hand panel of figure ~\ref{Y-La_60meV} 
shows $S(2\theta,\omega$) of YCoIn$_5$ and LaIrIn$_5$ for 60 meV incident energy  
for large scattering angles $\overline{2\theta}$ = 135$^{\circ}$ as black triangles (pointing up) and 
small scattering angles $\overline{2\theta}$ = 19${^\circ}$
as blue triangles (pointing down). All scattering of these non-magnetic samples
is of phonon origin so that a phonon scaling factor $R$ 
from large to small scattering angles can be determined. Empirically we find for both non-magnetic
samples that the high angle intensities scale to the low angle ones with a factor of 
$R$ = 1/8. The quality of this scaling is shown in the right hand panel of figure 
~\ref{Y-La_60meV}. The blue triangles (pointing down) are again the low angle scattering, now shown
on an expanded intensity scale and the black line is the high angle scattering scaled by  
$R$ = 1/8. While the overall intensity is well described, 
the low angle phonon scattering below 10 meV is under estimated by such a scaling 
and the fine structure of the low angle phonon scattering above 10 meV is not so well
reproduced. Method 2): next we check method 2 by scaling the two non-magnetic reference compounds to each other
with the averaged nuclear cross-section, again for grouped angles. The ratio of the averaged 
nuclear cross-sections of YCoIn$_5$ and LaIrIn$_5$ is 1.37. For large scattering angles the 60~meV data seem to 
scale rather well with this value; we find a scaling factor of 1.3. However, for small scattering angles
they do not. Here we find that the two non-magnetic data sets scale best 
with a factor of 1.1. This deviation is probably due to 
multiple scattering which seems stronger in the forward detectors. This finding makes 
this type of scaling some what arbitrary for the magnetic samples unless
a detailed phonon simulation and multiple scattering calculation is performed. Another obstacle 
of method 2) can be that often the non-magnetic reference samples are only measured at 
base temperature and the higher temperatures are obtained 
from Bose scaling. The bottom right panel of figure ~\ref{Y-La_60meV} shows that 
this can be traitorous: YCoIn$_5$ was measured at 5 (triangles) and 
250~K (circles). Scaling the scattering 
function $S(2\theta,\omega,T)$ at 5~K to 250~K with the Bose factor 
[$S(\theta,\omega,250K) = B(250K)/B(5) S(Q,\omega,5K)$ and 
$n(\omega,T)+1 = B(T) = 1/(1-exp(-h\omega/k_BT))$ ] gives the orange 
line which highly overestimates the actual scattering at 250~K. The Bose scaling does 
of course not take into account multiple phonon processes. 

We therefore discard the direct subtraction method for the phonon correction 
of the 60 meV data and rather apply the high to low $Q$ or large to small 
scattering angle $2\theta$ scaling.  However, for the
20 meV data which exhibit very little phonon scattering in the forward detectors 
(see left column of figure ~\ref{HETData}) the cross-section scaling looks fine and we 
correct the 20~meV data by subtracting the data of the cross-section scaled 
non-magnetic reference samples since the high angle data of the 20~meV data contain still a
non-negligible amount of magnetic scattering due to the smaller $Q$ values.

\section*{Acknowledgments}
The experiments at BESSY were supported by the BMBF through
project 05 ES3XBA/5. We thank L. Hamdan and the Cologne Mechanical
Workshop for skillful technical support. The wave function
density plots and transition matrix elements were calculated using the CrystalFieldTheory
package for Mathematica written by M. W. Haverkort. 

%\bibliography{references}

\begin{thebibliography}{41}
\expandafter\ifx\csname natexlab\endcsname\relax\def\natexlab#1{#1}\fi
\expandafter\ifx\csname bibnamefont\endcsname\relax
  \def\bibnamefont#1{#1}\fi
\expandafter\ifx\csname bibfnamefont\endcsname\relax
  \def\bibfnamefont#1{#1}\fi
\expandafter\ifx\csname citenamefont\endcsname\relax
  \def\citenamefont#1{#1}\fi
\expandafter\ifx\csname url\endcsname\relax
  \def\url#1{\texttt{#1}}\fi
\expandafter\ifx\csname urlprefix\endcsname\relax\def\urlprefix{URL }\fi
\providecommand{\bibinfo}[2]{#2}
\providecommand{\eprint}[2][]{\url{#2}}

\bibitem[{\citenamefont{Petrovic
  et~al.}(2001{\natexlab{a}})\citenamefont{Petrovic, Pagliuso, Hundley,
  Movshovich, Sarrao, Thompson, Fisk, and Monthoux}}]{PetrovicSCinCeCoIn5}
\bibinfo{author}{\bibfnamefont{C.}~\bibnamefont{Petrovic}},
  \bibinfo{author}{\bibfnamefont{P.~G.} \bibnamefont{Pagliuso}},
  \bibinfo{author}{\bibfnamefont{M.~F.} \bibnamefont{Hundley}},
  \bibinfo{author}{\bibfnamefont{R.}~\bibnamefont{Movshovich}},
  \bibinfo{author}{\bibfnamefont{J.~L.} \bibnamefont{Sarrao}},
  \bibinfo{author}{\bibfnamefont{J.~D.} \bibnamefont{Thompson}},
  \bibinfo{author}{\bibfnamefont{Z.}~\bibnamefont{Fisk}}, \bibnamefont{and}
  \bibinfo{author}{\bibfnamefont{P.}~\bibnamefont{Monthoux}},
  \bibinfo{journal}{J. Phys.: Condens. Matter} \textbf{\bibinfo{volume}{13}},
  \bibinfo{pages}{L337} (\bibinfo{year}{2001}{\natexlab{a}}).

\bibitem[{\citenamefont{Sidorov et~al.}(2002)\citenamefont{Sidorov, Nicklas,
  Pagliuso, Sarrao, Bang, Balatsky, and Thompson}}]{Sidorov2002}
\bibinfo{author}{\bibfnamefont{V.~A.} \bibnamefont{Sidorov}},
  \bibinfo{author}{\bibfnamefont{M.}~\bibnamefont{Nicklas}},
  \bibinfo{author}{\bibfnamefont{P.~G.} \bibnamefont{Pagliuso}},
  \bibinfo{author}{\bibfnamefont{J.~L.} \bibnamefont{Sarrao}},
  \bibinfo{author}{\bibfnamefont{Y.}~\bibnamefont{Bang}},
  \bibinfo{author}{\bibfnamefont{A.~V.} \bibnamefont{Balatsky}},
  \bibnamefont{and} \bibinfo{author}{\bibfnamefont{J.~D.}
  \bibnamefont{Thompson}}, \bibinfo{journal}{Phys. Rev. Lett.}
  \textbf{\bibinfo{volume}{89}}, \bibinfo{pages}{157004}
  (\bibinfo{year}{2002}).

\bibitem[{\citenamefont{Petrovic
  et~al.}(2001{\natexlab{b}})\citenamefont{Petrovic, Movshovich, Jaime,
  Pagliuso, Hundley, Sarrao, Fisk, and Thompson}}]{PetrovicSCinCeIrIn5}
\bibinfo{author}{\bibfnamefont{C.}~\bibnamefont{Petrovic}},
  \bibinfo{author}{\bibfnamefont{R.}~\bibnamefont{Movshovich}},
  \bibinfo{author}{\bibfnamefont{M.}~\bibnamefont{Jaime}},
  \bibinfo{author}{\bibfnamefont{P.}~\bibnamefont{Pagliuso}},
  \bibinfo{author}{\bibfnamefont{M.}~\bibnamefont{Hundley}},
  \bibinfo{author}{\bibfnamefont{J.}~\bibnamefont{Sarrao}},
  \bibinfo{author}{\bibfnamefont{Z.}~\bibnamefont{Fisk}}, \bibnamefont{and}
  \bibinfo{author}{\bibfnamefont{J.}~\bibnamefont{Thompson}},
  \bibinfo{journal}{Europhys. Lett.} \textbf{\bibinfo{volume}{53}},
  \bibinfo{pages}{354} (\bibinfo{year}{2001}{\natexlab{b}}).

\bibitem[{\citenamefont{Hegger et~al.}(2000)\citenamefont{Hegger, Petrovic,
  Moshopoulou, Hundley, Sarrao, Fisk, and Thompson}}]{HeggerSCinCeRhIn5}
\bibinfo{author}{\bibfnamefont{H.}~\bibnamefont{Hegger}},
  \bibinfo{author}{\bibfnamefont{C.}~\bibnamefont{Petrovic}},
  \bibinfo{author}{\bibfnamefont{E.~G.} \bibnamefont{Moshopoulou}},
  \bibinfo{author}{\bibfnamefont{M.~F.} \bibnamefont{Hundley}},
  \bibinfo{author}{\bibfnamefont{J.~L.} \bibnamefont{Sarrao}},
  \bibinfo{author}{\bibfnamefont{Z.}~\bibnamefont{Fisk}}, \bibnamefont{and}
  \bibinfo{author}{\bibfnamefont{J.~D.} \bibnamefont{Thompson}},
  \bibinfo{journal}{Phys. Rev. Lett.} \textbf{\bibinfo{volume}{84}},
  \bibinfo{pages}{4986} (\bibinfo{year}{2000}).

\bibitem[{\citenamefont{Knafo et~al.}(2003)\citenamefont{Knafo, Raymond, Fak,
  Lapertot, Canfield, and Flouquet}}]{Knafo2003}
\bibinfo{author}{\bibfnamefont{W.}~\bibnamefont{Knafo}},
  \bibinfo{author}{\bibfnamefont{S.}~\bibnamefont{Raymond}},
  \bibinfo{author}{\bibfnamefont{B.}~\bibnamefont{Fak}},
  \bibinfo{author}{\bibfnamefont{G.}~\bibnamefont{Lapertot}},
  \bibinfo{author}{\bibfnamefont{P.}~\bibnamefont{Canfield}}, \bibnamefont{and}
  \bibinfo{author}{\bibfnamefont{J.}~\bibnamefont{Flouquet}},
  \bibinfo{journal}{J. Phys.: Condens. Matter} \textbf{\bibinfo{volume}{15}},
  \bibinfo{pages}{3741} (\bibinfo{year}{2003}).

\bibitem[{\citenamefont{Monthoux and Lonzarich}(2002)}]{Monthoux2002}
\bibinfo{author}{\bibfnamefont{P.}~\bibnamefont{Monthoux}} \bibnamefont{and}
  \bibinfo{author}{\bibfnamefont{G.}~\bibnamefont{Lonzarich}},
  \bibinfo{journal}{Phys. Rev. B} \textbf{\bibinfo{volume}{66}},
  \bibinfo{pages}{224504} (\bibinfo{year}{2002}).

\bibitem[{\citenamefont{Thalmeier and Zwicknagl}(2005)}]{ThalmeierBook}
\bibinfo{author}{\bibfnamefont{P.}~\bibnamefont{Thalmeier}} \bibnamefont{and}
  \bibinfo{author}{\bibfnamefont{G.}~\bibnamefont{Zwicknagl}},
  \emph{\bibinfo{title}{Handbook on the Physics and Chemistry of Rare Earths}},
  vol.~\bibinfo{volume}{34} (\bibinfo{publisher}{Ed K.A. Gschneider, Jr.,
  J.-C.G. B\"unzli and V.K. Pecharsky}, \bibinfo{year}{2005}),
  \bibinfo{note}{and references therein}.

\bibitem[{\citenamefont{Nakatsuji et~al.}(2004)\citenamefont{Nakatsuji, Pines,
  and Fisk}}]{Nakatsuji2004}
\bibinfo{author}{\bibfnamefont{S.}~\bibnamefont{Nakatsuji}},
  \bibinfo{author}{\bibfnamefont{D.}~\bibnamefont{Pines}}, \bibnamefont{and}
  \bibinfo{author}{\bibfnamefont{Z.}~\bibnamefont{Fisk}},
  \bibinfo{journal}{Phys. Rev. Lett.} \textbf{\bibinfo{volume}{92}},
  \bibinfo{pages}{016401} (\bibinfo{year}{2004}).

\bibitem[{\citenamefont{Curro et~al.}(2004)\citenamefont{Curro, Young,
  Schmalian, and Pines}}]{Curro2004}
\bibinfo{author}{\bibfnamefont{N.~J.} \bibnamefont{Curro}},
  \bibinfo{author}{\bibfnamefont{B.-L.} \bibnamefont{Young}},
  \bibinfo{author}{\bibfnamefont{J.}~\bibnamefont{Schmalian}},
  \bibnamefont{and} \bibinfo{author}{\bibfnamefont{D.}~\bibnamefont{Pines}},
  \bibinfo{journal}{Phys. Rev. B} \textbf{\bibinfo{volume}{70}},
  \bibinfo{pages}{235117} (\bibinfo{year}{2004}).

\bibitem[{\citenamefont{Yang and Pines}(2008)}]{Yang2008}
\bibinfo{author}{\bibfnamefont{Y.-f.} \bibnamefont{Yang}} \bibnamefont{and}
  \bibinfo{author}{\bibfnamefont{D.}~\bibnamefont{Pines}},
  \bibinfo{journal}{Phys. Rev. Lett.} \textbf{\bibinfo{volume}{100}},
  \bibinfo{eid}{096404} (\bibinfo{year}{2008}).

\bibitem[{\citenamefont{Yang et~al.}(2008)\citenamefont{Yang, Fisk, Lee,
  Thompson, and Pines}}]{Yang2008a}
\bibinfo{author}{\bibfnamefont{Y.-f.} \bibnamefont{Yang}},
  \bibinfo{author}{\bibfnamefont{Z.}~\bibnamefont{Fisk}},
  \bibinfo{author}{\bibfnamefont{H.-O.} \bibnamefont{Lee}},
  \bibinfo{author}{\bibfnamefont{J.~D.} \bibnamefont{Thompson}},
  \bibnamefont{and} \bibinfo{author}{\bibfnamefont{D.}~\bibnamefont{Pines}},
  \bibinfo{journal}{Nature} \textbf{\bibinfo{volume}{454}},
  \bibinfo{pages}{611} (\bibinfo{year}{2008}).

\bibitem[{\citenamefont{Holland-Moritz
  et~al.}(1982)\citenamefont{Holland-Moritz, Wohlleben, and
  Loewenhaupt}}]{Holland-Moritz1982}
\bibinfo{author}{\bibfnamefont{E.}~\bibnamefont{Holland-Moritz}},
  \bibinfo{author}{\bibfnamefont{D.}~\bibnamefont{Wohlleben}},
  \bibnamefont{and}
  \bibinfo{author}{\bibfnamefont{M.}~\bibnamefont{Loewenhaupt}},
  \bibinfo{journal}{Phys. Rev. B} \textbf{\bibinfo{volume}{25}},
  \bibinfo{pages}{7482} (\bibinfo{year}{1982}).

\bibitem[{\citenamefont{Murani}(1983)}]{Murani1983}
\bibinfo{author}{\bibfnamefont{A.~P.}~\bibnamefont{Murani}}, \bibinfo{journal}{J.
  Phys. C} \textbf{\bibinfo{volume}{16}}, \bibinfo{pages}{6359}
  (\bibinfo{year}{1983}).

\bibitem[{\citenamefont{Severing
  et~al.}(1989{\natexlab{a}})\citenamefont{Severing, Holland-Moritz, and
  Frick}}]{Severing1989}
\bibinfo{author}{\bibfnamefont{A.}~\bibnamefont{Severing}},
  \bibinfo{author}{\bibfnamefont{E.}~\bibnamefont{Holland-Moritz}},
  \bibnamefont{and} \bibinfo{author}{\bibfnamefont{B.}~\bibnamefont{Frick}},
  \bibinfo{journal}{Phys. Rev. B} \textbf{\bibinfo{volume}{39}},
  \bibinfo{pages}{4164} (\bibinfo{year}{1989}{\natexlab{a}}).

\bibitem[{\citenamefont{Lawrence et~al.}(2001)\citenamefont{Lawrence,
  Riseborough, Booth, Sarrao, Thompson, and Osborn}}]{Lawrence2001}
\bibinfo{author}{\bibfnamefont{J.~M.} \bibnamefont{Lawrence}},
  \bibinfo{author}{\bibfnamefont{P.~S.} \bibnamefont{Riseborough}},
  \bibinfo{author}{\bibfnamefont{C.~H.} \bibnamefont{Booth}},
  \bibinfo{author}{\bibfnamefont{J.~L.} \bibnamefont{Sarrao}},
  \bibinfo{author}{\bibfnamefont{J.~D.} \bibnamefont{Thompson}},
  \bibnamefont{and} \bibinfo{author}{\bibfnamefont{R.}~\bibnamefont{Osborn}},
  \bibinfo{journal}{Phys. Rev. B} \textbf{\bibinfo{volume}{63}},
  \bibinfo{pages}{054427} (\bibinfo{year}{2001}).

\bibitem[{\citenamefont{Murani et~al.}(1996)\citenamefont{Murani, Severing, and
  Marshall}}]{Murani1996}
\bibinfo{author}{\bibfnamefont{A.~P.} \bibnamefont{Murani}},
  \bibinfo{author}{\bibfnamefont{A.}~\bibnamefont{Severing}}, \bibnamefont{and}
  \bibinfo{author}{\bibfnamefont{W.~G.} \bibnamefont{Marshall}},
  \bibinfo{journal}{Phys. Rev. B} \textbf{\bibinfo{volume}{53}},
  \bibinfo{pages}{2641} (\bibinfo{year}{1996}).

\bibitem[{\citenamefont{Bauer et~al.}(2004)\citenamefont{Bauer, Thompson,
  Sarrao, Morales, Wastin, Rebizant, Griveau, Javorsky, Boulet, Colineau
  et~al.}}]{BauerPRL2004}
\bibinfo{author}{\bibfnamefont{E.~D.} \bibnamefont{Bauer}},
  \bibinfo{author}{\bibfnamefont{J.~D.} \bibnamefont{Thompson}},
  \bibinfo{author}{\bibfnamefont{J.~L.} \bibnamefont{Sarrao}},
  \bibinfo{author}{\bibfnamefont{L.~A.} \bibnamefont{Morales}},
  \bibinfo{author}{\bibfnamefont{F.}~\bibnamefont{Wastin}},
  \bibinfo{author}{\bibfnamefont{J.}~\bibnamefont{Rebizant}},
  \bibinfo{author}{\bibfnamefont{J.~C.} \bibnamefont{Griveau}},
  \bibinfo{author}{\bibfnamefont{P.}~\bibnamefont{Javorsky}},
  \bibinfo{author}{\bibfnamefont{P.}~\bibnamefont{Boulet}},
  \bibinfo{author}{\bibfnamefont{E.}~\bibnamefont{Colineau}},
  \bibnamefont{et~al.}, \bibinfo{journal}{Phys. Rev. Lett.}
  \textbf{\bibinfo{volume}{93}}, \bibinfo{pages}{147005}
  (\bibinfo{year}{2004}).

\bibitem[{\citenamefont{Pagliuso et~al.}(2002)\citenamefont{Pagliuso, Curro,
  Moreno, Hundley, Thompson, Sarrao, and Fisk}}]{PagliusoPhysB320}
\bibinfo{author}{\bibfnamefont{P.}~\bibnamefont{Pagliuso}},
  \bibinfo{author}{\bibfnamefont{N.}~\bibnamefont{Curro}},
  \bibinfo{author}{\bibfnamefont{N.}~\bibnamefont{Moreno}},
  \bibinfo{author}{\bibfnamefont{M.}~\bibnamefont{Hundley}},
  \bibinfo{author}{\bibfnamefont{J.}~\bibnamefont{Thompson}},
  \bibinfo{author}{\bibfnamefont{J.}~\bibnamefont{Sarrao}}, \bibnamefont{and}
  \bibinfo{author}{\bibfnamefont{Z.}~\bibnamefont{Fisk}},
  \bibinfo{journal}{Physica B} \textbf{\bibinfo{volume}{320}},
  \bibinfo{pages}{370} (\bibinfo{year}{2002}).

\bibitem[{\citenamefont{Weber and Vojta}(2008)}]{WeberPRB77}
\bibinfo{author}{\bibfnamefont{H.}~\bibnamefont{Weber}} \bibnamefont{and}
  \bibinfo{author}{\bibfnamefont{M.}~\bibnamefont{Vojta}},
  \bibinfo{journal}{Phys. Rev. B} \textbf{\bibinfo{volume}{77}},
  \bibinfo{eid}{125118} (\bibinfo{year}{2008}).

\bibitem[{\citenamefont{Mena et~al.}(2005)\citenamefont{Mena, van~der Marel,
  and Sarrao}}]{mena2005}
\bibinfo{author}{\bibfnamefont{F.~P.} \bibnamefont{Mena}},
  \bibinfo{author}{\bibfnamefont{D.}~\bibnamefont{van~der Marel}},
  \bibnamefont{and} \bibinfo{author}{\bibfnamefont{J.~L.}
  \bibnamefont{Sarrao}}, \bibinfo{journal}{Phys. Rev. B}
  \textbf{\bibinfo{volume}{72}}, \bibinfo{eid}{045119} (\bibinfo{year}{2005}).

\bibitem[{\citenamefont{Ghaemi and Senthil}(2007)}]{GhaemiPRB75}
\bibinfo{author}{\bibfnamefont{P.}~\bibnamefont{Ghaemi}} \bibnamefont{and}
  \bibinfo{author}{\bibfnamefont{T.}~\bibnamefont{Senthil}},
  \bibinfo{journal}{Phys. Rev. B} \textbf{\bibinfo{volume}{75}},
  \bibinfo{eid}{144412} (\bibinfo{year}{2007}).

\bibitem[{\citenamefont{Burch et~al.}(2007)\citenamefont{Burch, Dordevic, Mena,
  Kuzmenko, van~der Marel, Sarrao, Jeffries, Bauer, Maple, and
  Basov}}]{BurchPRB75}
\bibinfo{author}{\bibfnamefont{K.~S.} \bibnamefont{Burch}},
  \bibinfo{author}{\bibfnamefont{S.~V.} \bibnamefont{Dordevic}},
  \bibinfo{author}{\bibfnamefont{F.~P.} \bibnamefont{Mena}},
  \bibinfo{author}{\bibfnamefont{A.~B.} \bibnamefont{Kuzmenko}},
  \bibinfo{author}{\bibfnamefont{D.}~\bibnamefont{van~der Marel}},
  \bibinfo{author}{\bibfnamefont{J.~L.} \bibnamefont{Sarrao}},
  \bibinfo{author}{\bibfnamefont{J.~R.} \bibnamefont{Jeffries}},
  \bibinfo{author}{\bibfnamefont{E.~D.} \bibnamefont{Bauer}},
  \bibinfo{author}{\bibfnamefont{M.~B.} \bibnamefont{Maple}}, \bibnamefont{and}
  \bibinfo{author}{\bibfnamefont{D.~N.} \bibnamefont{Basov}},
  \bibinfo{journal}{Phys. Rev. B} \textbf{\bibinfo{volume}{75}},
  \bibinfo{eid}{054523} (\bibinfo{year}{2007}).

\bibitem[{\citenamefont{Kubo and Hotta}(2006)}]{KuboJPSJ75}
\bibinfo{author}{\bibfnamefont{K.}~\bibnamefont{Kubo}} \bibnamefont{and}
  \bibinfo{author}{\bibfnamefont{T.}~\bibnamefont{Hotta}}, \bibinfo{journal}{J.
  Phys. Soc. Jpn.} \textbf{\bibinfo{volume}{75}}, \bibinfo{pages}{083702}
  (\bibinfo{year}{2006}).

\bibitem[{\citenamefont{Moreno and Coleman}(2000)}]{MorenoPRL84}
\bibinfo{author}{\bibfnamefont{J.}~\bibnamefont{Moreno}} \bibnamefont{and}
  \bibinfo{author}{\bibfnamefont{P.}~\bibnamefont{Coleman}},
  \bibinfo{journal}{Phys. Rev. Lett.} \textbf{\bibinfo{volume}{84}},
  \bibinfo{pages}{342} (\bibinfo{year}{2000}).

\bibitem[{\citenamefont{Curro et~al.}(2001)\citenamefont{Curro, Simovic,
  Hammel, Pagliuso, Sarrao, Thompson, and Martins}}]{CurroPRB64}
\bibinfo{author}{\bibfnamefont{N.~J.} \bibnamefont{Curro}},
  \bibinfo{author}{\bibfnamefont{B.}~\bibnamefont{Simovic}},
  \bibinfo{author}{\bibfnamefont{P.~C.} \bibnamefont{Hammel}},
  \bibinfo{author}{\bibfnamefont{P.~G.} \bibnamefont{Pagliuso}},
  \bibinfo{author}{\bibfnamefont{J.~L.} \bibnamefont{Sarrao}},
  \bibinfo{author}{\bibfnamefont{J.~D.} \bibnamefont{Thompson}},
  \bibnamefont{and} \bibinfo{author}{\bibfnamefont{G.~B.}
  \bibnamefont{Martins}}, \bibinfo{journal}{Phys. Rev. B}
  \textbf{\bibinfo{volume}{64}}, \bibinfo{pages}{180514}
  (\bibinfo{year}{2001}).

\bibitem[{\citenamefont{Takeuchi et~al.}(2001)\citenamefont{Takeuchi, Inoue,
  Sugiyama, Aoki, Tokiwa, Haga, Kindo, and \={O}nuki}}]{TakeuchiJPSJ70}
\bibinfo{author}{\bibfnamefont{T.}~\bibnamefont{Takeuchi}},
  \bibinfo{author}{\bibfnamefont{T.}~\bibnamefont{Inoue}},
  \bibinfo{author}{\bibfnamefont{K.}~\bibnamefont{Sugiyama}},
  \bibinfo{author}{\bibfnamefont{D.}~\bibnamefont{Aoki}},
  \bibinfo{author}{\bibfnamefont{Y.}~\bibnamefont{Tokiwa}},
  \bibinfo{author}{\bibfnamefont{Y.}~\bibnamefont{Haga}},
  \bibinfo{author}{\bibfnamefont{K.}~\bibnamefont{Kindo}}, \bibnamefont{and}
  \bibinfo{author}{\bibfnamefont{Y.}~\bibnamefont{\={O}nuki}},
  \bibinfo{journal}{J. Phys. Soc. Jpn.} \textbf{\bibinfo{volume}{70}},
  \bibinfo{pages}{877} (\bibinfo{year}{2001}).

\bibitem[{\citenamefont{Nakatsuji et~al.}(2002)\citenamefont{Nakatsuji, Yeo,
  Balicas, Fisk, Schlottmann, Pagliuso, Moreno, Sarrao, and
  Thompson}}]{NakatsujiPRL89}
\bibinfo{author}{\bibfnamefont{S.}~\bibnamefont{Nakatsuji}},
  \bibinfo{author}{\bibfnamefont{S.}~\bibnamefont{Yeo}},
  \bibinfo{author}{\bibfnamefont{L.}~\bibnamefont{Balicas}},
  \bibinfo{author}{\bibfnamefont{Z.}~\bibnamefont{Fisk}},
  \bibinfo{author}{\bibfnamefont{P.}~\bibnamefont{Schlottmann}},
  \bibinfo{author}{\bibfnamefont{P.}~\bibnamefont{Pagliuso}},
  \bibinfo{author}{\bibfnamefont{N.}~\bibnamefont{Moreno}},
  \bibinfo{author}{\bibfnamefont{J.}~\bibnamefont{Sarrao}}, \bibnamefont{and}
  \bibinfo{author}{\bibfnamefont{J.}~\bibnamefont{Thompson}},
  \bibinfo{journal}{Phys. Rev. Lett.} \textbf{\bibinfo{volume}{89}},
  \bibinfo{pages}{106402} (\bibinfo{year}{2002}).

\bibitem[{\citenamefont{Shishido et~al.}(2002)\citenamefont{Shishido, Settai,
  Aoki, Ikeda, Nakawaki, Nakamura, Iizuka, Inada, Sugiyama, Takeuchi
  et~al.}}]{ShishidoJPSJ71}
\bibinfo{author}{\bibfnamefont{H.}~\bibnamefont{Shishido}},
  \bibinfo{author}{\bibfnamefont{R.}~\bibnamefont{Settai}},
  \bibinfo{author}{\bibfnamefont{D.}~\bibnamefont{Aoki}},
  \bibinfo{author}{\bibfnamefont{S.}~\bibnamefont{Ikeda}},
  \bibinfo{author}{\bibfnamefont{H.}~\bibnamefont{Nakawaki}},
  \bibinfo{author}{\bibfnamefont{N.}~\bibnamefont{Nakamura}},
  \bibinfo{author}{\bibfnamefont{T.}~\bibnamefont{Iizuka}},
  \bibinfo{author}{\bibfnamefont{Y.}~\bibnamefont{Inada}},
  \bibinfo{author}{\bibfnamefont{K.}~\bibnamefont{Sugiyama}},
  \bibinfo{author}{\bibfnamefont{T.}~\bibnamefont{Takeuchi}},
  \bibnamefont{et~al.}, \bibinfo{journal}{J. Phys. Soc. Jpn.}
  \textbf{\bibinfo{volume}{71}}, \bibinfo{pages}{162} (\bibinfo{year}{2002}).

\bibitem[{\citenamefont{Christianson et~al.}(2004)\citenamefont{Christianson,
  Bauer, Lawrence, Riseborough, Moreno, Pagliuso, Sarrao, Thompson,
  Goremychkin, Trouw et~al.}}]{ChristiansonPRB70}
\bibinfo{author}{\bibfnamefont{A.~D.} \bibnamefont{Christianson}},
  \bibinfo{author}{\bibfnamefont{E.~D.} \bibnamefont{Bauer}},
  \bibinfo{author}{\bibfnamefont{J.~M.} \bibnamefont{Lawrence}},
  \bibinfo{author}{\bibfnamefont{P.~S.} \bibnamefont{Riseborough}},
  \bibinfo{author}{\bibfnamefont{N.~O.} \bibnamefont{Moreno}},
  \bibinfo{author}{\bibfnamefont{P.~G.} \bibnamefont{Pagliuso}},
  \bibinfo{author}{\bibfnamefont{J.~L.} \bibnamefont{Sarrao}},
  \bibinfo{author}{\bibfnamefont{J.~D.} \bibnamefont{Thompson}},
  \bibinfo{author}{\bibfnamefont{E.~A.} \bibnamefont{Goremychkin}},
  \bibinfo{author}{\bibfnamefont{F.~R.} \bibnamefont{Trouw}},
  \bibnamefont{et~al.}, \bibinfo{journal}{Phys. Rev. B}
  \textbf{\bibinfo{volume}{70}}, \bibinfo{pages}{134505}
  (\bibinfo{year}{2004}).

\bibitem[{\citenamefont{Christianson et~al.}(2002)\citenamefont{Christianson,
  Lawrence, Pagliuso, Moreno, Sarrao, Thompson, Riseborough, Kern, Goremychkin,
  and Lacerda}}]{ChristiansonPRB66}
\bibinfo{author}{\bibfnamefont{A.~D.} \bibnamefont{Christianson}},
  \bibinfo{author}{\bibfnamefont{J.~M.} \bibnamefont{Lawrence}},
  \bibinfo{author}{\bibfnamefont{P.~G.} \bibnamefont{Pagliuso}},
  \bibinfo{author}{\bibfnamefont{N.~O.} \bibnamefont{Moreno}},
  \bibinfo{author}{\bibfnamefont{J.~L.} \bibnamefont{Sarrao}},
  \bibinfo{author}{\bibfnamefont{J.~D.} \bibnamefont{Thompson}},
  \bibinfo{author}{\bibfnamefont{P.~S.} \bibnamefont{Riseborough}},
  \bibinfo{author}{\bibfnamefont{S.}~\bibnamefont{Kern}},
  \bibinfo{author}{\bibfnamefont{E.~A.} \bibnamefont{Goremychkin}},
  \bibnamefont{and} \bibinfo{author}{\bibfnamefont{A.~H.}
  \bibnamefont{Lacerda}}, \bibinfo{journal}{Phys. Rev. B}
  \textbf{\bibinfo{volume}{66}}, \bibinfo{pages}{193102}
  (\bibinfo{year}{2002}).

\bibitem[{\citenamefont{Hansmann et~al.}(2008)\citenamefont{Hansmann, Severing,
  Hu, Haverkort, Chang, Klein, Tanaka, Hsieh, Lin, Chen
  et~al.}}]{HansmannPRL100}
\bibinfo{author}{\bibfnamefont{P.}~\bibnamefont{Hansmann}},
  \bibinfo{author}{\bibfnamefont{A.}~\bibnamefont{Severing}},
  \bibinfo{author}{\bibfnamefont{Z.}~\bibnamefont{Hu}},
  \bibinfo{author}{\bibfnamefont{M.~W.} \bibnamefont{Haverkort}},
  \bibinfo{author}{\bibfnamefont{C.~F.} \bibnamefont{Chang}},
  \bibinfo{author}{\bibfnamefont{S.}~\bibnamefont{Klein}},
  \bibinfo{author}{\bibfnamefont{A.}~\bibnamefont{Tanaka}},
  \bibinfo{author}{\bibfnamefont{H.~H.} \bibnamefont{Hsieh}},
  \bibinfo{author}{\bibfnamefont{H.-J.} \bibnamefont{Lin}},
  \bibinfo{author}{\bibfnamefont{C.~T.} \bibnamefont{Chen}},
  \bibnamefont{et~al.}, \bibinfo{journal}{Phys. Rev. Lett.}
  \textbf{\bibinfo{volume}{100}}, \bibinfo{eid}{066405} (\bibinfo{year}{2008}).

\bibitem[{\citenamefont{Willers et~al.}(2009)\citenamefont{Willers, k,
  Hollmann, K\"{o}rner, Hu, Tanaka, Schmitz, Enderle, Lapertot, Tjeng
  et~al.}}]{WillersPRB80}
\bibinfo{author}{\bibfnamefont{T.}~\bibnamefont{Willers}},
  \bibinfo{author}{\bibfnamefont{B.~F.} \bibnamefont{k}},
  \bibinfo{author}{\bibfnamefont{N.}~\bibnamefont{Hollmann}},
  \bibinfo{author}{\bibfnamefont{P.~O.} \bibnamefont{K\"{o}rner}},
  \bibinfo{author}{\bibfnamefont{Z.}~\bibnamefont{Hu}},
  \bibinfo{author}{\bibfnamefont{A.}~\bibnamefont{Tanaka}},
  \bibinfo{author}{\bibfnamefont{D.}~\bibnamefont{Schmitz}},
  \bibinfo{author}{\bibfnamefont{M.}~\bibnamefont{Enderle}},
  \bibinfo{author}{\bibfnamefont{G.}~\bibnamefont{Lapertot}},
  \bibinfo{author}{\bibfnamefont{L.~H.} \bibnamefont{Tjeng}},
  \bibnamefont{et~al.}, \bibinfo{journal}{Phys. Rev. B}
  \textbf{\bibinfo{volume}{80}}, \bibinfo{eid}{115106} (\bibinfo{year}{2009}).

\bibitem[{\citenamefont{Tanaka and Jo}(1994)}]{TanakaJPSC63}
\bibinfo{author}{\bibfnamefont{A.}~\bibnamefont{Tanaka}} \bibnamefont{and}
  \bibinfo{author}{\bibfnamefont{T.}~\bibnamefont{Jo}}, \bibinfo{journal}{J.
  Phys. Soc. Jpn.} \textbf{\bibinfo{volume}{63}}, \bibinfo{pages}{2788}
  (\bibinfo{year}{1994}).

\bibitem[{\citenamefont{Severing
  et~al.}(1989{\natexlab{b}})\citenamefont{Severing, Holland-Moritz, Rainford,
  Culverhouse, and Frick}}]{Severing1989a}
\bibinfo{author}{\bibfnamefont{A.}~\bibnamefont{Severing}},
  \bibinfo{author}{\bibfnamefont{E.}~\bibnamefont{Holland-Moritz}},
  \bibinfo{author}{\bibfnamefont{B.~D.} \bibnamefont{Rainford}},
  \bibinfo{author}{\bibfnamefont{S.~R.} \bibnamefont{Culverhouse}},
  \bibnamefont{and} \bibinfo{author}{\bibfnamefont{B.}~\bibnamefont{Frick}},
  \bibinfo{journal}{Phys. Rev. B} \textbf{\bibinfo{volume}{39}},
  \bibinfo{pages}{2557} (\bibinfo{year}{1989}{\natexlab{b}}).

\bibitem[{\citenamefont{Willers}(2007)}]{mythesis}
\bibinfo{author}{\bibfnamefont{T.}~\bibnamefont{Willers}},
  \bibinfo{journal}{Diploma Thesis,University of Cologne}
  (\bibinfo{year}{2007}).

\bibitem[{\citenamefont{Fulde and Loewenhaupt}(1985)}]{Fulde1986}
\bibinfo{author}{\bibfnamefont{P.}~\bibnamefont{Fulde}} \bibnamefont{and}
  \bibinfo{author}{\bibfnamefont{M.}~\bibnamefont{Loewenhaupt}},
  \bibinfo{journal}{Advances in Physics} \textbf{\bibinfo{volume}{34}},
  \bibinfo{pages}{589} (\bibinfo{year}{1985}).

\bibitem[{\citenamefont{Stock et~al.}(2008)\citenamefont{Stock, Broholm, Hudis,
  Kang, and Petrovic}}]{Stock2008}
\bibinfo{author}{\bibfnamefont{C.}~\bibnamefont{Stock}},
  \bibinfo{author}{\bibfnamefont{C.}~\bibnamefont{Broholm}},
  \bibinfo{author}{\bibfnamefont{J.}~\bibnamefont{Hudis}},
  \bibinfo{author}{\bibfnamefont{H.~J.} \bibnamefont{Kang}}, \bibnamefont{and}
  \bibinfo{author}{\bibfnamefont{C.}~\bibnamefont{Petrovic}},
  \bibinfo{journal}{Phys. Rev. Lett.} \textbf{\bibinfo{volume}{100}},
  \bibinfo{eid}{087001} (\bibinfo{year}{2008}).

\bibitem[{\citenamefont{Horn et~al.}(1981)\citenamefont{Horn, Holland-Moritz,
  Loewenhaupt, Steglich, Scheuer, Benoit, and Flouquet}}]{horn1981}
\bibinfo{author}{\bibfnamefont{S.}~\bibnamefont{Horn}},
  \bibinfo{author}{\bibfnamefont{E.}~\bibnamefont{Holland-Moritz}},
  \bibinfo{author}{\bibfnamefont{M.}~\bibnamefont{Loewenhaupt}},
  \bibinfo{author}{\bibfnamefont{F.}~\bibnamefont{Steglich}},
  \bibinfo{author}{\bibfnamefont{H.}~\bibnamefont{Scheuer}},
  \bibinfo{author}{\bibfnamefont{A.}~\bibnamefont{Benoit}}, \bibnamefont{and}
  \bibinfo{author}{\bibfnamefont{J.}~\bibnamefont{Flouquet}},
  \bibinfo{journal}{Phys. Rev. B} \textbf{\bibinfo{volume}{23}},
  \bibinfo{pages}{3171} (\bibinfo{year}{1981}).

\bibitem[{\citenamefont{Grosche et~al.}(1996)\citenamefont{Grosche, Julian,
  Mathur, and Lonzarich}}]{Grosche1996}
\bibinfo{author}{\bibfnamefont{F.}~\bibnamefont{Grosche}},
  \bibinfo{author}{\bibfnamefont{S.}~\bibnamefont{Julian}},
  \bibinfo{author}{\bibfnamefont{N.}~\bibnamefont{Mathur}}, \bibnamefont{and}
  \bibinfo{author}{\bibfnamefont{G.}~\bibnamefont{Lonzarich}},
  \bibinfo{journal}{Physica B} \textbf{\bibinfo{volume}{224}},
  \bibinfo{pages}{50} (\bibinfo{year}{1996}).

\bibitem[{\citenamefont{Mathur et~al.}(1998)\citenamefont{Mathur, Grosche,
  Julian, Walker, Freye, Haselwimmer, and Lonzarich}}]{MathurNature394}
\bibinfo{author}{\bibfnamefont{N.~D.} \bibnamefont{Mathur}},
  \bibinfo{author}{\bibfnamefont{F.~M.} \bibnamefont{Grosche}},
  \bibinfo{author}{\bibfnamefont{S.~R.} \bibnamefont{Julian}},
  \bibinfo{author}{\bibfnamefont{I.~R.} \bibnamefont{Walker}},
  \bibinfo{author}{\bibfnamefont{D.~M.} \bibnamefont{Freye}},
  \bibinfo{author}{\bibfnamefont{R.~K.~W.} \bibnamefont{Haselwimmer}},
  \bibnamefont{and} \bibinfo{author}{\bibfnamefont{G.~G.}
  \bibnamefont{Lonzarich}}, \bibinfo{journal}{Nature}
  \textbf{\bibinfo{volume}{394}}, \bibinfo{pages}{39} (\bibinfo{year}{1998}).

\bibitem[{\citenamefont{Link and Jaccard}(1997)}]{link1997}
\bibinfo{author}{\bibfnamefont{P.}~\bibnamefont{Link}} \bibnamefont{and}
  \bibinfo{author}{\bibfnamefont{D.}~\bibnamefont{Jaccard}},
  \bibinfo{journal}{Physica B} \textbf{\bibinfo{volume}{230}},
  \bibinfo{pages}{31} (\bibinfo{year}{1997}).

\end{thebibliography}

\end{document}